\documentclass[a4paper,fleqn]{cas-sc}

\usepackage[numbers]{natbib}

\usepackage{amsmath}
\usepackage{longtable}
\DeclareMathOperator*{\minimize}{minimize}

\usepackage{xcolor}
\definecolor{revisionblue}{HTML}{1F4E79}

\def\tsc#1{\csdef{#1}{\textsc{\lowercase{#1}}\xspace}}
\tsc{WGM}
\tsc{QE}
\tsc{EP}
\tsc{PMS}
\tsc{BEC}
\tsc{DE}

\begin{document}
\let\WriteBookmarks\relax
\def\floatpagepagefraction{1}
\def\textpagefraction{.001}
\shorttitle{Variable-lattice-density optimization of pin-fin heat sinks under high-Reynolds-number flow conditions}
\shortauthors{Y. Furusawa et~al.}

\title [mode = title]{Variable-lattice-density optimization of pin-fin heat sinks under high-Reynolds-number flow conditions}                      

\author[1]{Yoshikatsu Furusawa}[auid=000,bioid=1,
                        orcid=0000-0002-6271-0097]
\cormark[1]
\ead{yoshikatsu.furusawa@nature-architects.com}

\credit{Writing -- original draft, Visualization, Validation, Software, Methodology, Investigation, Formal analysis, Data curation, Conceptualization}

\affiliation[1]{organization={Nature Architects, Inc.},
                addressline={1-3-8 Nihonbashi-Ningyocho}, 
                city={Chuo},
                postcode={103-0013}, 
                state={Tokyo},
                country={Japan}}

\author[1]{Kunitaka Shintani}[orcid=0009-0007-7185-1127]
\author[1]{Shunsuke Hirotani}[orcid=0009-0008-9467-4749]

\credit{Writing -- review \& editing, Methodology}

\author[2]{Kentaro Yaji}[orcid=0000-0002-4309-2043]
\affiliation[2]{organization={The University of Osaka},
                addressline={2-1 Yamadaoka},
                city={Suita}, 
                postcode={565-0871}, 
                postcodesep={}, 
                state={Osaka},
                country={Japan}}

\cortext[cor1]{Corresponding author}

\begin{abstract}
This study extends variable-lattice-density optimization to high-Reynolds-number flows for the design of periodically arranged pin-fin heat sinks. The effective permeability and Forchheimer coefficient are identified from unit-cell flow analyses of cylindrical pin-fin arrays and incorporated into a reduced model based on the Darcy--Forchheimer law for macroscopic design exploration. To enable stable optimization under high-Reynolds-number conditions, a dual-mesh framework is introduced, in which the flow field and sensitivities are evaluated on a fine mesh, whereas the design variables are updated on a coarse mesh corresponding to the unit-cell arrangement. For the baseline condition, geometry-resolved Reynolds-averaged Navier--Stokes (RANS) analysis of the reconstructed design gives a reduction in the root-mean-square deviation of temperature from its area-weighted mean from 4.27 K to 0.942 K, while the reduced model predicts a reduction from 4.00 K to 0.901 K. Additional calculations with higher inlet velocities and modified outlet locations show that the proposed method captures the dominant flow-redistribution trends under different operating and geometric conditions, although agreement in the thermal objective deteriorates when the geometry-resolved RANS results exhibit pronounced intra-solid temperature variations. These results indicate that the proposed approach is useful as a practical design-exploration method for identifying candidate pin-fin heat sink configurations under high-Reynolds-number conditions.

\end{abstract}

\begin{keywords}
Variable-lattice-density optimization \sep High-Reynolds-number flow \sep Forced convection \sep Heat sink
\end{keywords}

\maketitle
\hypersetup{
  pdfauthor={Yoshikatsu Furusawa; Kunitaka Shintani; Shunsuke Hirotani; Kentaro Yaji}
}
\section*{Nomenclature}
\begingroup
\small
\renewcommand{\LTcaptype}{}
\setlength{\LTleft}{0pt}
\setlength{\LTright}{\fill}
\setlength{\tabcolsep}{4pt}
\renewcommand{\arraystretch}{1.05}
\begin{longtable}{@{}p{0.23\linewidth}p{\dimexpr0.77\linewidth-8pt\relax}@{}}
\toprule
\textbf{Symbol} & \textbf{Description and unit} \\
\midrule
\endfirsthead
\multicolumn{2}{@{}l}{\textit{Nomenclature (continued)}} \\
\toprule
\textbf{Symbol} & \textbf{Description and unit} \\
\midrule
\endhead
\midrule
\multicolumn{2}{r@{}}{\textit{Continued on next page}} \\
\endfoot
\bottomrule
\endlastfoot
\multicolumn{2}{@{}l}{\textbf{Roman symbols}} \\*
$c_{p,f}$ & Fluid specific heat at constant pressure (J\,kg$^{-1}$\,K$^{-1}$) \\
$d$ & Design variable determining the cylinder radius (--) \\
$g,g_p$ & Normalized porosity and pressure-drop constraints, respectively (--) \\
$J$ & Root-mean-square temperature deviation from $\bar T$ (K) \\
$k$ & Turbulent kinetic energy (m$^2$\,s$^{-2}$) \\
$k_f,k_s$ & Fluid and solid thermal conductivities (W\,m$^{-1}$\,K$^{-1}$) \\
$k_0$ & Stagnant effective thermal conductivity (W\,m$^{-1}$\,K$^{-1}$) \\
$k_{\mathrm{eff}}$ & Apparent effective thermal conductivity (W\,m$^{-1}$\,K$^{-1}$) \\
$p$ & Pressure (Pa) \\
$\dot q$ & Volumetric heat-generation rate (W\,m$^{-3}$) \\
$r$ & Cylinder radius (m) \\
$Re_\kappa$ & Permeability-based Reynolds number (--) \\
$s$ & Unit-cell size (m) \\
$T,\bar T$ & Temperature and its area-weighted mean over $\Omega$ (K) \\
$\mathbf u$ & Velocity vector: Darcy velocity in the reduced model; local fluid velocity in unaveraged geometry-resolved field plots (m\,s$^{-1}$) \\
$u$ & Darcy velocity in the $x$ direction used for unit-cell coefficient identification (m\,s$^{-1}$) \\
$\mathbf u^{\mathrm{cell}}$ & Geometry-resolved fluid velocity in the unit cell (m\,s$^{-1}$) \\
$U_{\mathrm{in}}$ & Prescribed inlet velocity (m\,s$^{-1}$) \\
\addlinespace
\multicolumn{2}{@{}l}{\textbf{Greek symbols}} \\*
$\beta$ & Forchheimer coefficient (m$^{-1}$) \\
$\Delta p$ & Pressure drop (Pa) \\
$\varepsilon$ & Turbulent kinetic-energy dissipation rate (m$^2$\,s$^{-3}$) \\
$\kappa$ & Permeability (m$^2$) \\
$\mu,\mu_t$ & Molecular and turbulent dynamic viscosities (Pa\,s) \\
$\rho_f$ & Fluid density (kg\,m$^{-3}$) \\
$\phi$ & Local porosity (--) \\
$\bar\phi^{\max}$ & Upper limit on the area-weighted average porosity (--) \\
$\Omega$ & Macroscopic design domain \\
\addlinespace
\multicolumn{2}{@{}l}{\textbf{Abbreviations}} \\*
DF & Darcy--Forchheimer \\
GR & Geometry-resolved \\
MMA & Method of Moving Asymptotes \\
RANS & Reynolds-averaged Navier--Stokes \\
RM & Reduced model \\
\end{longtable}
\endgroup

\section{Introduction}

Forced convection at high Reynolds numbers is a key phenomenon governing the performance of a wide range of thermal engineering systems, including thermal management systems for vehicles and aircraft, turbomachinery, and high-performance electronic devices~\cite{togun25,mudawar00,yeranee21,marshall19}. In such systems, thermal management structures must not only remove heat within a limited installation volume but also suppress spatial temperature nonuniformity, which can lead to local hot spots and large thermal gradients. In addition, the design must satisfy manufacturability and cost requirements simultaneously. However, because both flow and heat transfer exhibit strong nonlinearity at high Reynolds numbers, it is not straightforward to systematically search a broad design space for high-performance geometries.

Heat sinks composed of periodically arranged pin fins have long been used in such applications because they can achieve high heat-transfer performance through increased heat-transfer area. In classical studies, the effects of geometric parameters such as pin diameter, pitch, aspect ratio, and array configuration have been investigated in detail for both inline and staggered pin-fin arrays through experiments and numerical simulations~\cite{sparrow80,vanfossen82,metzger82,metzger84,armstrong88}. More recent numerical studies have examined the effects of pin-fin dimensions and arrangements on heat transfer and pressure drop, providing a basis for evaluating pin-fin array configurations. Zhao \emph{et al.}~\cite{zhao26} numerically investigated locally concentrated pin-fin arrangements and optimized pin diameters and pitches to maximize heat transfer and minimize pressure drop. Building on such geometric design variables, Mihalko and Basak~\cite{mihalko25} applied Bayesian optimization with transfer learning to explore heterogeneous pin-fin arrays by maximizing a thermal--hydraulic efficiency parameter. The computational cost of repeated flow and heat-transfer analyses has also motivated the development of surrogate-model-based optimization. Ghosh \emph{et al.}~\cite{ghosh24} employed a multi-fidelity approach to optimize pin-fin arrays under a pressure-drop constraint. These developments highlight the importance of efficient evaluation of flow and heat transfer in pin-fin design exploration.

As a mathematical design methodology for automatically exploring a broad design space, topology optimization, originally developed in structural mechanics~\cite{bendse88}, has also been applied to fluid mechanics and heat-sink design~\cite{borrvallxx,yoon10,pietropaoli19,subramaniam19,kobayashi21,alexandersen20,fawaz22}. In topology optimization, the design domain is generally discretized into many elements, and design variables such as material density are assigned to each element. Then, a conjugate heat-transfer problem is solved while optimizing the material distribution (i.e., the channel geometry) under temperature-related objectives or constraints. Although this framework is powerful in that it can automatically generate high-performance flow paths that would be difficult for humans to conceive, two major barriers remain for its direct use in practical design. First, the resulting optimized shapes tend to contain numerous fine branched channels and complex free-form surfaces, making them difficult to mass-produce using conventional manufacturing processes such as machining, pressing, and casting~\cite{fawaz22}. Metal additive manufacturing can be an effective option for fabricating complex geometries, but it has significant disadvantages for mass production because of cost and production-time limitations. Second, although topology optimization for turbulent flows has also been investigated~\cite{alexandersen20,dilgen18,zhao21,sun23,bayat26}, the computational costs remain large, particularly because of wall-resolution requirements. To address these issues, one possible strategy is to optimize the size distribution of representative features in structures composed of periodically repeated similar shapes.

A promising framework for reducing computational cost is to homogenize the periodic structure and describe the entire design domain using its effective properties~\cite{bendse88,banthiya22,thillaithevan25,padhy24,takezawa19,takezawa24,saito26,kikuchi26,ohtani26,yanagihara26}. In this approach, detailed flow and heat-transfer analyses are performed for a unit cell, and the spatially averaged behavior is formulated in terms of effective properties. The entire design domain can then be represented as a homogeneous porous medium without explicitly resolving the detailed geometry. If local porosity or lattice density is introduced as the design variable, material distribution optimization can be carried out at a practical computational cost while reproducing macroscopic quantities such as pressure drop and temperature distribution with reasonable accuracy, compared to geometry-resolved simulations. Based on this idea, Takezawa \emph{et al.}~\cite{takezawa19,takezawa24} proposed variable-lattice-density optimization, in which the local dimension of a periodic lattice is treated as the design variable, and demonstrated a design methodology that controls flow and heat-transfer characteristics in the design domain using a Brinkman--Forchheimer-type porous flow model. In this context, Banthiya \emph{et al.}~\cite{banthiya22} linked design variables to duct dimensions through homogenization and optimized the distribution of internal duct sizes in a heat-conduction problem without explicitly solving convection in the ducts. Thillaithevan \emph{et al.}~\cite{thillaithevan25} proposed a subspace method in which momentum is modeled using a multiscale flow model based on cell permeability, while heat transfer is modeled as a cell-wise advection-diffusion problem with consistency of temperature and heat flux enforced between adjacent unit cells; this approach yielded the optimized radius distribution of periodically arranged pin fins and the corresponding temperature field under forced convection. Padhy \emph{et al.}~\cite{padhy24} proposed a method in which the selection weights of multiple microstructure candidates (cross-sectional shapes), together with their sizes and orientations, are represented and optimized as continuous fields using a neural network with spatial coordinates as inputs, thereby obtaining flow channels that minimize pressure drop. Saito \emph{et al.}~\cite{saito26} proposed a multiscale topology optimization method for heat sinks containing both void and graded-lattice regions using a two-layer Darcy--Forchheimer model. Kikuchi \emph{et al.}~\cite{kikuchi26} applied a homogenization-based topology optimization framework to non-uniform strut-based lattice heat sinks and demonstrated that the optimized lattice-density distribution and thermal--hydraulic performance depend on the choice of unit-cell geometry and the heat-sink aspect ratio. Ohtani \emph{et al.}~\cite{ohtani26} developed an effective porous-media model for gyroid two-fluid heat exchangers and optimized the spatial distribution of the wall thickness of the triply periodic minimal surface (TPMS). Yanagihara \emph{et al.}~\cite{yanagihara26} optimized the isosurface-threshold distribution of a Primitive TPMS two-fluid heat exchanger using a Brinkman--Forchheimer-based macroscopic model and validated the optimized design through detailed-geometry simulations and experiments. These recent studies demonstrate the growing applicability of homogenization-based graded-lattice optimization to heat sinks and two-fluid heat exchangers. Although variable-lattice-density optimization does not explicitly restrict the internal geometry of the unit cell, choosing a simple pin-fin geometry makes it possible to express the design as a periodically arranged cylindrical pin-fin array. Compared with the complex structures typically obtained by topology optimization, such a design is much more compatible with existing manufacturing processes such as machining, casting, and press forming, and therefore offers a substantial advantage in terms of mass producibility.

Previously reported homogenization-based design methods for heat sinks have mainly considered laminar flow regimes. 
In practical devices handling high heat fluxes, the Reynolds number is often much larger. Under such conditions, inertial effects and turbulent transport influence flow and heat transfer. Accordingly, a design framework is needed that extends variable-lattice-density optimization to high-Reynolds-number flows while restricting the final geometry to mass-producible periodic pin fins. In variable-lattice-density optimization, the pin-fin region is represented as a porous medium; therefore, near-wall mesh refinement around individual fins is not required during optimization. In addition, a turbulence model for flow in a porous medium has been proposed~\cite{nakayama08}, which makes it possible to perform design exploration while accounting for the turbulent flow in the porous medium.

Based on the above considerations, the objective of this study is to extend variable-lattice-density optimization based on Darcy--Forchheimer-type homogenization to high-Reynolds-number flows and to establish a design exploration method for heat sinks composed of periodically arranged pin-fin arrays, using idealized two-dimensional model problems. Specifically, (1) the Reynolds-averaged Navier--Stokes (RANS) equations are solved with a turbulence model for a unit cell containing cylindrical pin fins to identify the effective properties of the porous flow model; (2) the pin-fin radius is introduced as the design variable, and a reduced model using the Darcy--Forchheimer law is constructed using the effective properties obtained in step (1); and (3) promising pin-fin radius distributions are explored using the reduced model, with the area-normalized $L_2$ norm of the temperature deviation from the area-averaged temperature as the objective function. The reduced model is evaluated against RANS simulations of the reconstructed pin-fin arrangements in which the individual pin-fin geometries are explicitly resolved (geometry-resolved RANS). These simulations serve as the numerical reference for assessing agreement in predicted fields and objective values, the relative ranking of candidate designs, and improvements in temperature uniformity.

The remainder of this paper is organized as follows. Section~2 presents the formulation of variable-lattice-density optimization for high-Reynolds-number flows, including the identification of effective properties based on unit-cell analysis, the reduced model based on the porous approximation, and the formulation of the optimization problem. Section~3 first assesses the proposed method by comparing the reduced model with geometry-resolved RANS analyses for randomly generated design variable distributions, and then presents the optimized design obtained under the baseline condition with its flow and heat-transfer characteristics. An additional case with a pressure-drop constraint is also considered, followed by comparisons with simplified flow models and a smaller unit cell. The effects of inlet velocity and outlet position on the optimized design and predictive accuracy are also examined. Finally, Section~4 summarizes the conclusions of this study.

\section{Methods}

This study optimizes the radius distribution of periodically arranged cylindrical pin fins in a two-dimensional design domain $\Omega \subset \mathbb{R}^2$. The proposed method consists of two stages: (1) construction of a homogenization-based porous approximation model, and (2) gradient-based optimization using that model. In stage~(1), the flow in a unit cell containing a cylinder is solved, and the effective properties of the porous model, namely the permeability $\kappa$ and the Forchheimer coefficient $\beta$, are identified. In stage~(2), these effective properties are assigned to the macroscopic design domain as functions of the cylinder radius, and the radius distribution that minimizes the objective function is obtained by gradient-based optimization.

\subsection{Governing equations}

In the present study, the entire design domain is regarded as an isotropic porous medium, and the flow around the pin fins is approximated as porous flow. To describe the macroscopic turbulent flow, we used the momentum equation for porous flow derived by Nakayama and Kuwahara~\cite{nakayama99}. In this formulation, Darcy and Forchheimer terms are included as well as turbulent momentum diffusion through the eddy viscosity:
\begin{equation}
\frac{\rho_f}{\phi}(\mathbf{u}\cdot\nabla)\frac{\mathbf{u}}{\phi}
=
-\nabla p
+
\nabla \cdot \left[\frac{\mu+\mu_t}{\phi}(\nabla\mathbf{u}+(\nabla\mathbf{u})^{\mathsf{T}})\right]
-
\frac{\mu}{\kappa}\mathbf{u}
-
\rho_f\beta |\mathbf{u}| \mathbf{u}.
\label{eq:ns}
\end{equation}
Here, $\mathbf{u}$ is the Darcy velocity vector, $\rho_f$ is the fluid density, $\phi$ is the porosity, $p$ is the pressure, $\mu$ is the dynamic viscosity, $\mu_t$ is the turbulent viscosity, and the superscript $\mathsf{T}$ denotes the transpose. This equation extends the Brinkman--Forchheimer equation~\cite{nield17} to turbulent flows by modeling turbulent diffusion using the turbulent viscosity. The effective permeability $\kappa$ and Forchheimer coefficient $\beta$ appearing in these terms are identified from the Darcy--Forchheimer pressure-drop relation~\cite{nield17}:
\begin{equation}
\nabla p = -\frac{\mu}{\kappa}\mathbf{u} - \rho_f \beta |\mathbf{u}| \mathbf{u}.
\label{eq:df}
\end{equation}
Kuwahara \emph{et al.}~\cite{kuwahara98} showed that this relation remains valid for periodically arranged square rods even in the high-Reynolds-number regime. Based on this finding, the present study uses this relation to model pressure drop in pin-fin arrays.

Whereas $\beta$ may depend on the flow direction, it is treated as isotropic as a practical approximation for design exploration. The direction-dependent formulation examined in Appendix~\ref{app:directional_drag} did not consistently improve the objective values of the optimized designs evaluated using geometry-resolved RANS across the tested inlet velocities. This comparison does not establish that inertial anisotropy is negligible.

The governing equations of the present study consist of Eq.~\eqref{eq:ns}, together with the continuity equation and the energy equation:
\begin{equation}
\nabla \cdot \mathbf{u} = 0,
\label{eq:continuity}
\end{equation}
\begin{equation}
\rho_f c_{p,f} \mathbf{u}\cdot\nabla T
= \nabla\cdot(k_{\mathrm{eff}}\nabla T)
+
\dot{q}.
\label{eq:energy}
\end{equation}
In Eq.~\eqref{eq:energy}, $c_{p,f}$ is the specific heat of the fluid at constant pressure, $T$ is the temperature, $k_{\mathrm{eff}}$ is the effective thermal conductivity, and $\dot{q}$ is the source term. In the present study, $k_{\mathrm{eff}}$ is modeled as an isotropic apparent conductivity consisting of the stagnant conductivity identified from unit-cell conduction analyses and the turbulent heat-diffusion contribution
\begin{equation}
k_{\mathrm{eff}} = k_0(d) + \phi\frac{c_{p,f}\mu_t}{Pr_t},
\label{eq:keff}
\end{equation}
where $k_0(d)$ is the stagnant effective conductivity of the actual fluid--solid unit cell and $Pr_t = 0.9$ is the turbulent Prandtl number. Using $k_0(d)$ in Eq. (5) approximates the conductive contribution under flow by that obtained from the stagnant unit-cell problem. At high Reynolds numbers, the additional heat transport associated with turbulent mixing is represented by the eddy-diffusivity term, providing a flow-dependent thermal closure for the single macroscopic temperature field. These equations enable the calculation of fluid flow and heat transfer without resolving the detailed flow structures within the porous medium, and therefore act as a reduced model for design exploration.

An additional optimization incorporating thermal dispersion while retaining the same $k_0(d)$ is also presented in Appendix~\ref{app:thermal_dispersion}. The optimized pin-fin radius distribution was almost unchanged from that obtained with Eq.~\eqref{eq:keff} under the baseline condition. The simpler thermal model without an explicit thermal-dispersion term is therefore retained for the design exploration reported in the main text.

The turbulent viscosity $\mu_t$ is given by the Nakayama--Kuwahara model~\cite{nakayama08}, which extends the standard $k$--$\varepsilon$ turbulence model~\cite{launder74} to porous flows:
\begin{equation}
\mu_t = \rho_f C_{\mu}\frac{k^2}{\varepsilon\phi},
\label{eq:mut}
\end{equation}
\begin{equation}
\rho_f\mathbf{u}\cdot\nabla\left(\frac{k}{\phi}\right)
=
P_k + P_{k,\mathrm{pm}}
+
\nabla \cdot \left[\left(\mu+\frac{\mu_t}{\sigma_k}\right)\nabla k\right]
-
\rho_f\varepsilon,
\label{eq:k}
\end{equation}
\begin{equation}
\rho_f\mathbf{u}\cdot\nabla\left(\frac{\varepsilon}{\phi}\right)
=
P_{\varepsilon} + P_{\varepsilon,\mathrm{pm}} - D_{\varepsilon}
+
\nabla \cdot \left[\left(\mu+\frac{\mu_t}{\sigma_{\varepsilon}}\right)\nabla \varepsilon\right].
\label{eq:eps}
\end{equation}
Here, $P_k$ denotes the production of turbulent kinetic energy by the mean velocity gradients, and $\rho_f\varepsilon$ represents its dissipation. The production and destruction terms in the $\varepsilon$-transport equation are defined as
\begin{equation}
P_{\varepsilon}=C_{\varepsilon1}\frac{\varepsilon}{k}P_k,
\qquad
D_{\varepsilon}=C_{\varepsilon2}\rho_f\frac{\varepsilon^2}{k}.
\label{eq:eps_terms}
\end{equation}
The standard $k$--$\varepsilon$ model constants are $C_\mu=0.09$, $C_{\varepsilon1}=1.44$, $C_{\varepsilon2}=1.92$, $\sigma_k=1.0$, and $\sigma_\varepsilon=1.3$~\cite{launder74}. The additional production terms $P_{k,\mathrm{pm}}$ and $P_{\varepsilon,\mathrm{pm}}$ follow the Nakayama--Kuwahara model~\cite{nakayama08}. The standard wall function is used at rigid-wall boundaries.

For the unit-cell analysis, the governing equations are the continuity equation and the following two-dimensional steady RANS equation with an eddy viscosity model:
\begin{equation}
\rho_f (\mathbf{u}^{\mathrm{cell}}\cdot\nabla)\mathbf{u}^{\mathrm{cell}}
=
-\nabla p^{\mathrm{cell}}
+
\nabla \cdot \left[(\mu+\mu_t^{\mathrm{cell}})\left(\nabla\mathbf{u}^{\mathrm{cell}}+(\nabla\mathbf{u}^{\mathrm{cell}})^{\mathsf{T}}\right)\right].
\label{eq:ge_unitcell}
\end{equation}
Here, the superscript $\mathrm{cell}$ denotes the local fields resolved within the fluid region of the unit cell. In the unit-cell RANS computations, the standard $k$--$\varepsilon$ turbulence model is used, i.e., Eqs.~\eqref{eq:k} and \eqref{eq:eps} with $\phi=1$ and $P_{k,\mathrm{pm}}=P_{\varepsilon,\mathrm{pm}}=0$.

\subsection{Estimation of effective properties}
\label{sec:effective_properties}

\begin{figure}
	\centering
\includegraphics[width=.95\textwidth]{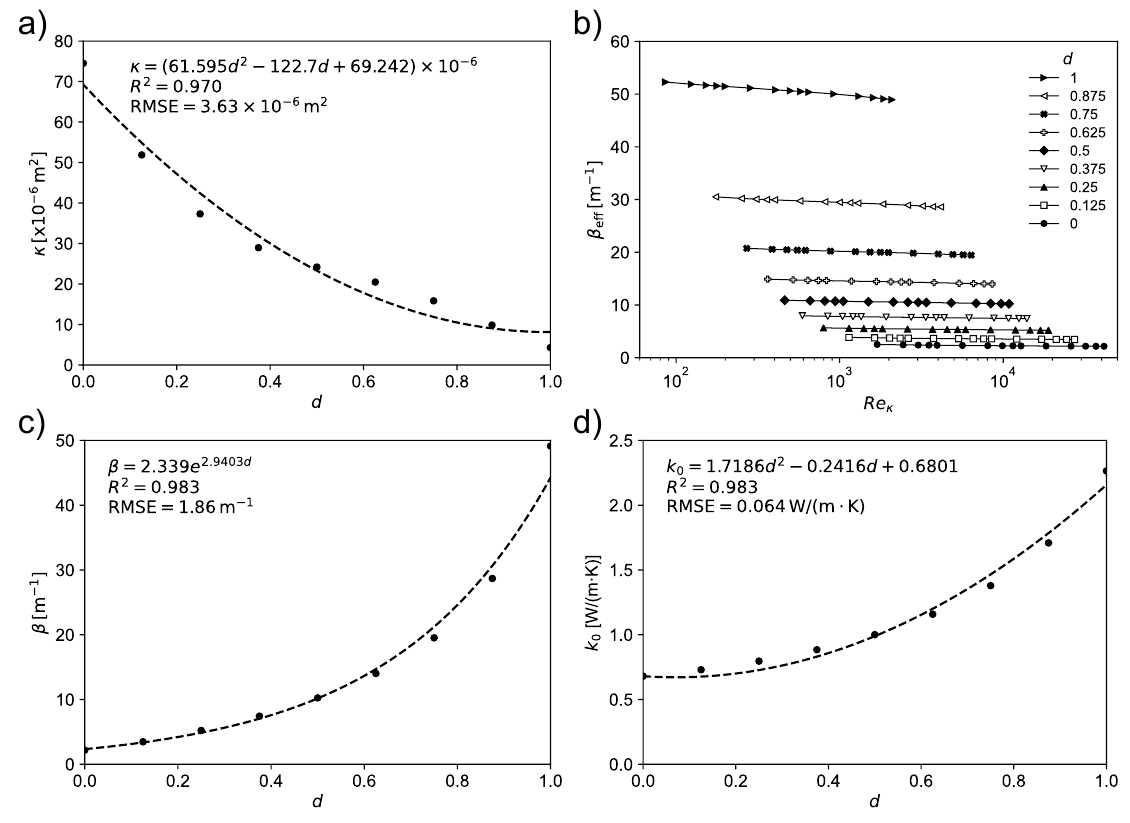}
\caption{Unit-cell effective properties: (a) permeability $\kappa$ versus $d$; (b) pointwise effective Forchheimer coefficient $\beta_{\mathrm{eff}}$ versus permeability-based Reynolds number $Re_\kappa$; (c) fitted Forchheimer coefficient $\beta$ versus $d$; and (d) stagnant effective thermal conductivity $k_0$ versus $d$. In (b), $\beta_{\mathrm{eff}}=[\Delta p/s-(\mu/\kappa)u]/(\rho_f u^2)$ and $Re_\kappa=\rho_f u\sqrt{\kappa}/\mu$ are evaluated for each nonzero-flow result using the independently identified $\kappa$; lines connect the data points. RMSEs denote the root-mean-square errors of the correlations relative to the identified unit-cell values.}
	\label{kappabeta}
\end{figure}

The effective properties $\kappa$ and $\beta$ are estimated using a unit-cell-based numerical homogenization approach as in Refs.~\cite{takezawa19,takezawa24}. The effective properties of a periodically structured object are determined from a unit cell and then used to represent the macroscale body as a homogeneous continuum. In variable-lattice-density optimization, objects are placed in a unit cell $Y=[0,s]\times[0,s]$, and multiple simulations are carried out while varying the object size and inlet conditions to obtain the effective properties of the unit cell. The resulting relation between the Darcy velocity (hereinafter referred to simply as velocity unless otherwise noted) and the pressure drop $\Delta p$ is then fitted sequentially to the linear and quadratic terms of the Darcy--Forchheimer law, Eq.~\eqref{eq:df}, by the least-squares method to extract $\kappa$ and $\beta$. The resulting $\kappa$ and $\beta$ are treated as isotropic scalars, subsequently fitted as functions of the design variable, as shown in Fig.~\ref{kappabeta}, and then used in Eq.~\eqref{eq:ns}. The design variable $d(\mathbf{x})\in[0,1]$ is defined to correspond one-to-one to the object size in the unit cell.

\begin{figure}
	\centering
	\includegraphics[width=.6\textwidth]{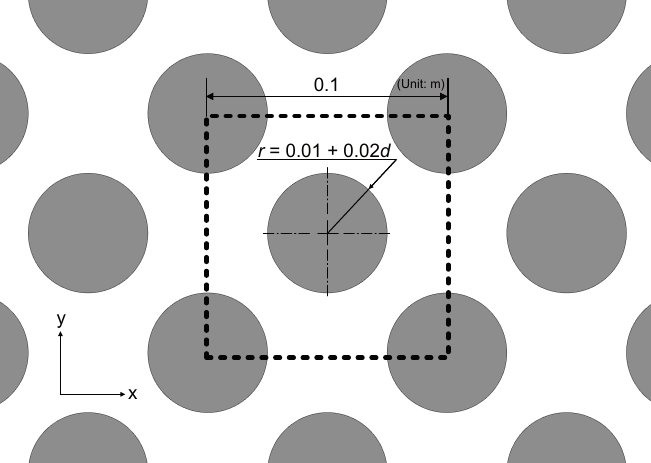}
	\caption{Schematic of the unit cell and its dimensions.}
	\label{unitcell}
\end{figure}

In the present study, the unit cell is a square with a side length of $s=0.1$~m as shown in Fig.~\ref{unitcell}. Cylinders of radius $r(d)=(0.01+0.02d)$~m are arranged in a staggered square lattice at the cell center and corners. Accordingly, the radius $r(d)$ also corresponds one-to-one to the local porosity $\phi(d)$:
\begin{equation}
\phi(d)=1-\frac{2\pi r(d)^2}{s^2}.
\label{eq:porosity}
\end{equation}
The pitch $s$ is fixed, and only the radius distribution is optimized. Although the unit cell may appear too large for the usual homogenization assumption, previous studies have shown that even for such large periodic structures, macroscopic characteristics can be approximated with sufficient accuracy~\cite{takezawa24}.

The permeability $\kappa$ and the Forchheimer coefficient $\beta$ were identified sequentially to separate the estimation of the linear viscous term from that of the nonlinear inertial term. For each prescribed design variable $d$, steady unit-cell simulations were first performed without a turbulence model, using pressure differences $\Delta p$ ranging from 0 to 0.01~Pa at intervals of 0.001~Pa. The pressure drop was imposed in the $x$ direction using the same periodic and no-slip boundary conditions as in the identification of $\beta$ described below. The permeability was estimated by the least-squares method from the resulting relationship between the Darcy velocity and the imposed pressure gradient using
\begin{equation}
\frac{\Delta p}{s} = \frac{\mu}{\kappa}u.
\label{eq:permeability_identification}
\end{equation}
The fourfold symmetry of the unit cell implies an isotropic in-plane permeability in the Darcy limit.

The resulting $\kappa(d)$ was then held fixed while $\beta(d)$ was identified from the unit-cell RANS simulations. On the two faces normal to the $x$-axis in Fig.~\protect\ref{unitcell}, periodic boundary conditions with prescribed pressure differences $\Delta p$ of 0--50, 100--500, and 1000--5000~Pa, with intervals of 10, 100, and 1000~Pa, respectively, are imposed, whereas on the two faces normal to the $y$-axis, periodic boundary conditions with $\Delta p=0$~Pa are used. The cylinder walls are treated as no-slip walls. Thus, the scalar coefficient was identified using only the results for a driving orientation of $0^\circ$ relative to the $x$-axis. For each unit-cell result $n$, the Darcy velocity was evaluated as
\begin{equation}
\mathbf{u}_n
=\frac{1}{|Y|}\int_{Y_f}\mathbf{u}^{\mathrm{cell}}_n\,dY,
\label{eq:cell_darcy_velocity}
\end{equation}
where $Y_f$ is the fluid region of the unit cell. The linear Darcy contribution was subtracted from the imposed pressure-drop gradient, and the remaining contribution was fitted by the least-squares method using
\begin{equation}
\frac{\Delta p}{s}-\frac{\mu}{\kappa(d)}u
=\rho_f\beta(d)|u|u,
\label{eq:beta_identification}
\end{equation}
where $u$ is the Darcy velocity in the $x$ direction.

In addition, the stagnant effective thermal conductivity $k_0(d)$ was evaluated by solving steady heat conduction in the fluid and solid regions of the same unit cell, with no flow or internal heat source. A macroscopic temperature gradient $\Delta T/s$ was imposed in the $x$ direction using periodic temperature fluctuations. Temperature and normal heat flux were continuous at the fluid--solid interface. The effective conductivity was calculated from the cell-averaged conductive heat flux as
\begin{equation}
k_0(d)=-\frac{s\langle q_x\rangle_Y}{\Delta T},
\label{eq:stagnant_conductivity}
\end{equation}
where $\langle q_x\rangle_Y$ denotes the conductive heat flux averaged over both the fluid and solid regions of the unit cell. The fourfold symmetry of the unit cell yields an isotropic in-plane conductivity.

Figure~\ref{kappabeta} summarizes the identified properties: $\kappa(d)$ in (a), the Reynolds-number dependence of the pointwise $\beta_{\mathrm{eff}}$ in (b), $\beta(d)$ in (c), and $k_0(d)$ in (d).

\subsection{Problem settings}

\begin{figure}
	\centering
	\includegraphics[width=.6\textwidth]{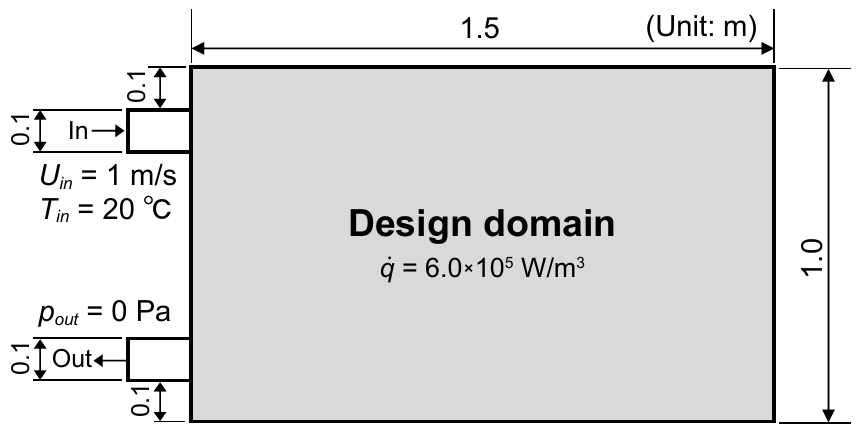}
	\caption{Computational domain and operating conditions.}
	\label{compdomain1}
\end{figure}

As shown in Fig.~\protect\ref{compdomain1}, the design domain $\Omega$ is a rectangle of size 1.5~m $\times$ 1.0~m. At the inlet $\Gamma_{\mathrm{in}}$, the velocity and temperature are fixed at $U_{\mathrm{in}}=1$~m/s and $T_{\mathrm{in}}=20\,^{\circ}\mathrm{C}$; the corresponding maximum Reynolds number based on permeability is $Re_{\kappa}=\rho_f|\mathbf{u}|\sqrt\kappa/\mu=10,857$. At the outlet $\Gamma_{\mathrm{out}}$, the pressure is fixed at $p_{\mathrm{out}}=0$~Pa. A uniform volumetric heat source $\dot{q}=6.0\times10^5$~W/m$^3$ is applied to both the fluid and solid regions within the design domain. Adiabatic no-slip conditions are applied to the walls. The thermophysical properties of the solid and fluid are taken as those of aluminum and water, respectively, as listed in Table~\ref{props}. These properties are held constant in the unit-cell, reduced-model, and geometry-resolved calculations to assess the proposed framework under consistent reference conditions. The design domain is divided into $15\times 10$ unit cells, and cylinders are placed in a staggered arrangement at the center and four corners of each unit cell, as described above.

\begin{table}[width=.9\linewidth,cols=3,pos=h]
\centering
\caption{Thermophysical properties used in the computations.}
\begin{tabular*}{\tblwidth}{@{} LLLL@{} }
\toprule
Property & Fluid (water) & Solid (aluminum) \\
\midrule
Density [kg/m$^3$] & 1000 & 2700 \\
Specific heat [J/(kg$\cdot$K)] & 4200 & 900 \\
Thermal conductivity [W/(m$\cdot$K)] & 0.6 & 238 \\
Dynamic viscosity [Pa$\cdot$s] & 0.001 & -- \\
\bottomrule
\end{tabular*}
\label{props}
\end{table}

The objective function is defined as the area-normalized $L_2$ norm of the temperature deviation from the area-weighted average temperature in the design domain, and the optimization problem is formulated as
\begin{equation}
\minimize_d \; J
=
\left(
\frac{1}{|\Omega|}
\int_{\Omega}
\left(T(\mathbf{x})-\bar{T}\right)^2
\,d\Omega
\right)^{1/2}
\label{eq:objective}
\end{equation}
subject to
\begin{equation}
g =
\frac{
\displaystyle \frac{1}{|\Omega|}
\int_{\Omega} \phi(d)\,d\Omega
}{
\bar{\phi}^{\max}
}
\le 1,
\label{eq:porosity_constraint}
\end{equation}
\begin{equation}
0 \le d(\mathbf{x}) \le 1
\qquad \text{in } \Omega.
\label{eq:design_bounds}
\end{equation}
Here, $\bar{T}$ is the area-weighted average temperature in the design domain. This objective corresponds to the root-mean-square temperature deviation from the area-weighted average temperature and is therefore used to quantify spatial temperature nonuniformity in the design domain. The maximum temperature and pressure drop are additionally evaluated using geometry-resolved RANS to characterize the thermal and hydraulic performance of the resulting designs. The parameter $\bar{\phi}^{\max}$ denotes the upper limit of the area-weighted average porosity, which is set to 0.7 in the present study. This bound is imposed as a global lattice-density constraint. Because smaller design variable values correspond to smaller pin-fin radii and therefore higher porosity, unconstrained optimization can generate excessively open, low-resistance regions over a large part of the domain. In such cases, the macroscopic flow distribution becomes highly sensitive to small local changes in the design variable because of the flow nonlinearity. The constraint is therefore used to avoid high-porosity layouts and to enable robust design exploration.

In the additional example presented in Section~\ref{sec:pressure_constraint}, the pressure drop between the inlet and outlet is also constrained as follows:
\begin{equation}
g_p(d)=\frac{\Delta p(d)}{\Delta p_{\mathrm{lim}}}\le1.
\label{eq:pressure_constraint}
\end{equation}
Here, $\Delta p(d)$ denotes the pressure drop between $\Gamma_{\mathrm{in}}$ and $\Gamma_{\mathrm{out}}$, and $\Delta p_{\mathrm{lim}}$ is the prescribed limit on the pressure drop, which is set to 1100~Pa in Section~\ref{sec:pressure_constraint}. This additional constraint is applied only to that example; the temperature-nonuniformity objective and the porosity constraint are retained.

\begin{figure}
	\centering
	\includegraphics[width=.6\textwidth]{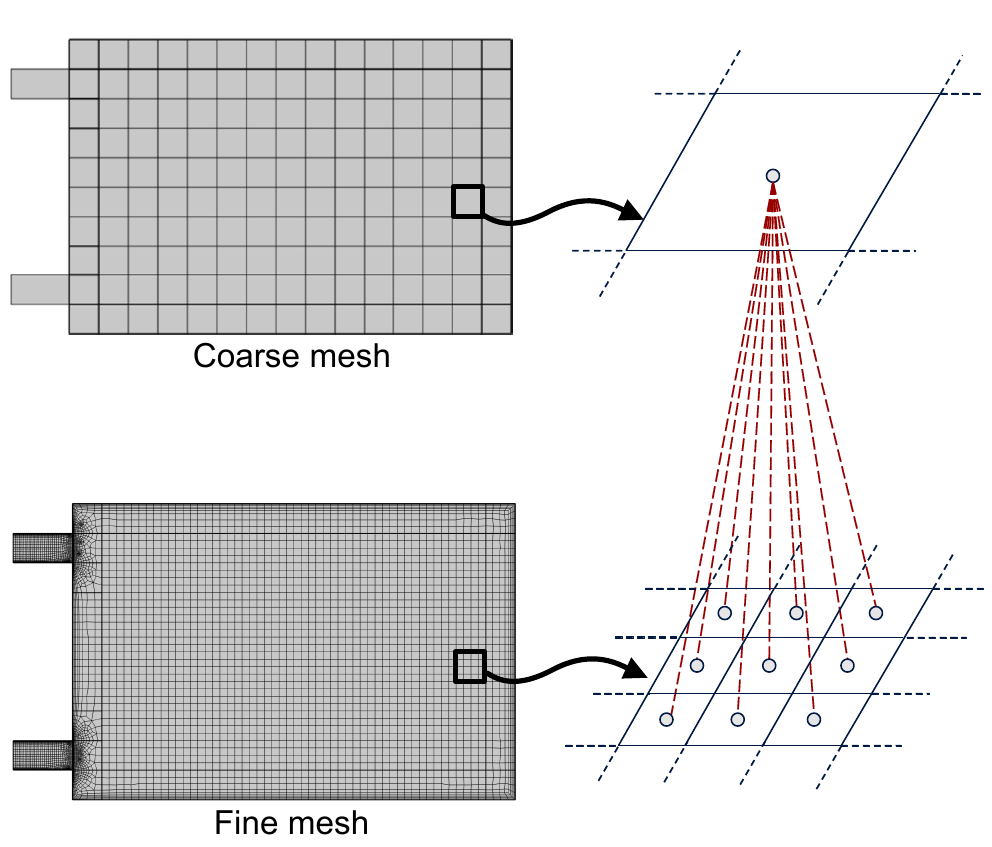}
	\caption{Coarse and fine meshes used in the computations.}
	\label{mesh}
\end{figure}

\subsection{Computational methods}

Because the pin fins are arranged regularly, a mesh composed of $15\times 10$ square elements is used so that the size of one computational cell corresponds to one unit cell. However, in order to treat high-Reynolds-number flow accurately, a sufficiently fine mesh that resolves the boundary layer is indispensable. Therefore, two meshes are employed as shown in Fig.~\protect\ref{mesh}. These two meshes are combined in the optimization procedure; the design variables are updated on the coarse mesh, while the governing equations are solved on the fine mesh. One optimization cycle is carried out in the following order:
\begin{enumerate}
\item Map the design variables from the coarse mesh to the fine mesh.
\item Solve the flow field on the fine mesh.
\item Compute the sensitivities on the fine mesh.
\item Sum the fine-element sensitivity contributions to obtain the sensitivities with respect to the coarse-mesh design variables.
\item Update the design variables on the coarse mesh.
\end{enumerate}
By repeating this sequence, the radius distribution of the pin fins is optimized. The coarse mesh consists of 150 elements, whereas the fine mesh (RM Mesh~2) consists of 6312 elements and includes fine elements near the walls. The sensitivity of the reduced-model results to the fine-mesh resolution is examined in Appendix~\ref{app:rm_mesh}. Because walls neither form nor disappear during optimization in this setting, the boundary-layer mesh is fixed and does not require updating, which enables stable optimization under high-Reynolds-number conditions.

In the present study, all computations are performed using COMSOL Multiphysics 6.3. The governing equations, Eqs.~\eqref{eq:ns}--\eqref{eq:energy}, \eqref{eq:k}, \eqref{eq:eps}, and \eqref{eq:ge_unitcell}, are solved by the finite element method. Pressure, velocity, and temperature are discretized using first-order Lagrange elements with stabilization based on the Galerkin least-squares method~\cite{hughes89}, while the design variable is defined elementwise. The relative solver tolerance was set to $10^{-3}$ for both the unit-cell and reduced-model calculations. Neither filtering nor projection is applied to the design variables. The sensitivities are computed using COMSOL's discrete adjoint method with automatic differentiation, and the design variables are updated using the Method of Moving Asymptotes (MMA)~\cite{svanbergxx} with a move limit of 0.1. The adjoint sensitivity calculation accounts for the dependence of the mean-flow, turbulence, and temperature equations on the design variables, including variations in the turbulence variables $k$ and $\varepsilon$ and the resulting changes in the eddy viscosity. The objective and constraint sensitivities with respect to a representative coarse-mesh design variable were verified against central finite differences, with maximum relative discrepancies below $1.8\times10^{-3}$ and $10^{-8}$ for the objective and constraint sensitivities, respectively, using the optimization solver settings. The design variable was initialized uniformly as $d=0.593$ throughout the design domain, giving a feasible reference design that satisfies the porosity constraint with $g\approx1.0$. The optimization was terminated after 100 iterations, which was found to be sufficient in preliminary tests to obtain converged optimized design variables. The final design was taken from the 100th iteration, at which the objective value and design variable distribution had already reached a nearly stationary state.

To assess how accurately the reduced model reproduces the underlying RANS predictions, the reconstructed pin-fin geometries were analyzed using geometry-resolved RANS simulations. The two-dimensional steady RANS equations were solved using the standard $k$--$\varepsilon$ turbulence model with wall functions on the pin-fin surfaces and channel walls. This model was adopted for consistency with the unit-cell calculations and for computational efficiency in evaluating multiple designs, although the predictive capability of separated and recirculating flows is subject to the limitations of the eddy-viscosity and wall-function approximations. In the present study, geometry-resolved RANS is used as a numerical reference for evaluating the reduced model. The relative tolerance of the steady-state solver was set to $10^{-5}$ to obtain converged temperature distributions. The same heat-source distribution and definition of the objective function were used in both the reduced and geometry-resolved models. In the geometry-resolved simulations, $J$ was evaluated before the unit-cell averaging used for comparing macroscopic field distributions.

\section{Results and Discussion}

\subsection{Numerical assessment of the proposed method for design exploration}
\label{sec:validation}

Since the performance of the reduced model is assessed by comparison with geometry-resolved RANS simulations in the following sections, the mesh convergence of these simulations is first examined. The mesh-sensitivity study used the reconstructed pin-fin geometry of the optimized design for the baseline case presented in Section~\ref{sec:baseline}. Five meshes, denoted GR Meshes~1--5 in order of increasing element count and containing 103,034--485,711 elements, were compared at $U_{\mathrm{in}}=1$~m/s. Table~\ref{tab:gr_mesh_sensitivity} summarizes the temperature-nonuniformity objective $J$ and the pressure drop $\Delta p$ between the inlet and outlet. The relative differences between GR Meshes~4 and~5, normalized by the GR Mesh~5 values, were 0.54\% for $J$ and 0.77\% for $\Delta p$. These results indicate that both quantities are sufficiently converged with respect to the number of elements for assessing the performance of the reduced model. Accordingly, meshes with approximately the same number of elements as GR Mesh~4 were used for the subsequent geometry-resolved calculations. For GR Mesh~4, the wall $y^+$ values ranged from 11.06, which is COMSOL's lower limit, to 30.55.

\begin{table}
\centering
\caption{Mesh sensitivity study for the geometry-resolved RANS calculation at $U_{\mathrm{in}}=1$~m/s. Relative differences are absolute differences from the GR Mesh~5 results, normalized by those results and expressed as percentages.}
\label{tab:gr_mesh_sensitivity}
\begin{tabular}{lrrrrr}
\toprule
Mesh & Elements & $J$ [K] & $\Delta p$ [Pa]
     & \multicolumn{2}{c}{Relative difference [\%]} \\
\cmidrule(lr){5-6}
     &          &         &
     & $J$ & $\Delta p$ \\
\midrule
GR Mesh~1 & 103,034 & 0.83979 & 1687.8 & 10.35 & 38.05 \\
GR Mesh~2 & 163,675 & 0.89474 & 1402.9 &  4.49 & 14.75 \\
GR Mesh~3 & 252,981 & 0.93005 & 1273.5 &  0.72 &  4.16 \\
GR Mesh~4 & 366,429 & 0.94179 & 1213.2 &  0.54 &  0.77 \\
GR Mesh~5 & 485,711 & 0.93676 & 1222.6 &    -- &    -- \\
\bottomrule
\end{tabular}
\end{table}

To assess the proposed method as a reduced model for design exploration, the results obtained from the model are then compared with those from geometry-resolved RANS analyses. Candidate design variable distributions with values in $[0, 1]$ were generated using a uniform random number generator. Only distributions satisfying the porosity constraint ($g\le1$) were retained, and this procedure was repeated until thirty feasible distributions were obtained. These distributions were then converted into cylinder arrays by dehomogenization, discretized with meshes containing approximately the same number of elements as GR Mesh~4, and analyzed by geometry-resolved RANS simulations for comparison with the reduced model results. In the dehomogenization process, the radii of the cylinders located at the corners of the unit cells were determined from the average of the design variables of the adjacent unit cells in order to suppress geometric discontinuities at cell boundaries.

\begin{figure}
	\centering
	\includegraphics[width=1.\textwidth]{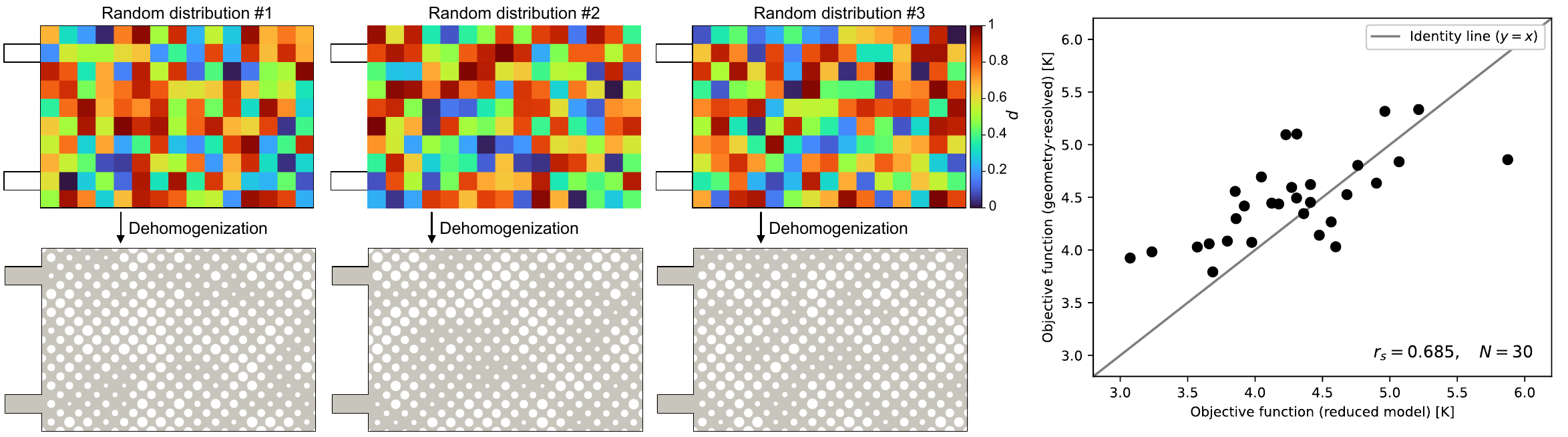}
	\caption{Examples of randomly generated design variable distributions and cylinder arrays reconstructed by dehomogenization, and a comparison of the objective values obtained from the reduced model and the geometry-resolved simulations. The solid line in the scatter plot denotes the identity line, corresponding to exact agreement between the two models.}
	\label{validation}
\end{figure}

Figure~\protect\ref{validation} shows examples of the generated design variable distributions, the corresponding reconstructed cylinder arrays, and a comparison of the objective function values obtained from the reduced model and from the geometry-resolved simulations. Although the objective values do not agree well, a certain correspondence is observed in the relative ranking of the thirty distributions. Specifically, Spearman's rank correlation coefficient calculated from these data,
\begin{equation}
r_s = 1 - \frac{6\sum_{n=1}^{N} D_n^2}{N^3-N},
\end{equation}
was $r_s=\mathrm{0.685}$. Here, $D_n$ denotes the difference between the ranks assigned to the objective value of each distribution by the reduced model and the geometry-resolved analysis, and $N$ denotes the number of samples ($N = 30$). The mean difference between the objective values predicted by the geometry-resolved and reduced models was 0.196~K, indicating that the reduced model underpredicted the objective value on average over this sample. These results indicate a moderate positive correlation in the relative ranking of candidate designs, although the reduced model does not necessarily reproduce the objective values with high quantitative accuracy.

For the optimized baseline geometry, the objective values obtained with GR Meshes~4 and~5 differed by 0.54\% (Table~\ref{tab:gr_mesh_sensitivity}). In the higher-velocity and modified-outlet cases discussed below, the geometry-resolved simulations show high-temperature regions within the cylinders and in recirculation regions, respectively, that are not fully captured by the reduced model (Figs.~\ref{uin125} and~\ref{veltemp12}). These observations indicate that the representation of local temperature variations affects the accuracy of the predicted objective value.

\subsection{Baseline optimization and geometry-resolved assessment}
\label{sec:baseline}

\begin{figure}
	\centering
	\includegraphics[width=1.\textwidth]{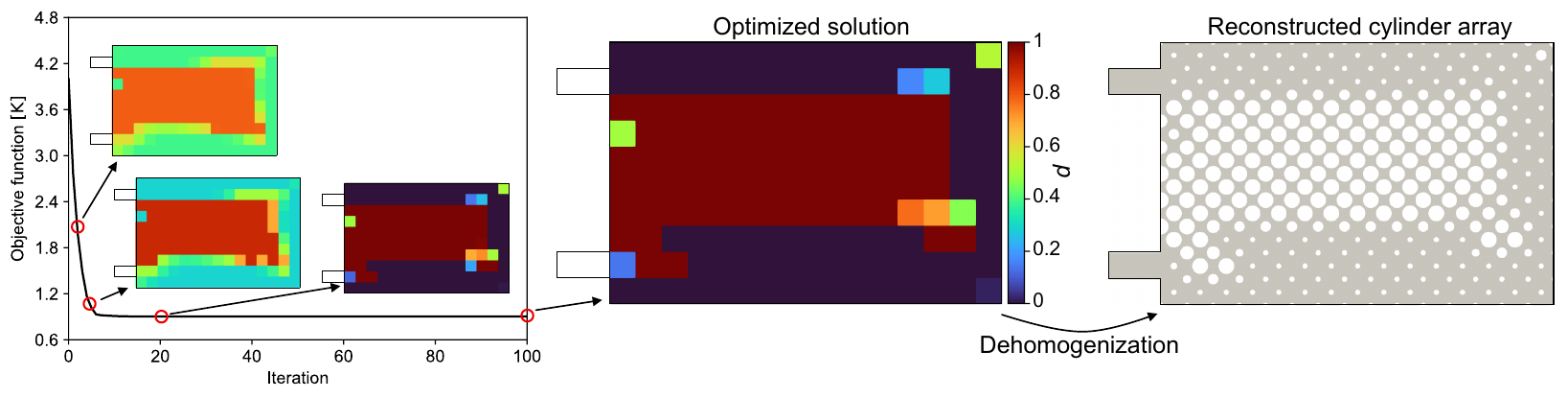}
	\caption{History of the objective value and the design variable distribution, and the cylinder array obtained from the optimization result.}
	\label{history}
\end{figure}

Whereas the previous subsection examined the rank consistency of the reduced model for randomly generated designs, the present subsection focuses on whether the design obtained by the proposed optimization retains its performance after reconstruction into an explicit pin-fin array and re-evaluation by geometry-resolved RANS simulation. Figure~\protect\ref{history} shows the convergence history of the objective value and snapshots of the design variable distribution. As shown in the figure, the optimization converged sufficiently within 20 steps, after which the objective value and the design variable distribution did not change significantly. The objective value decreased from 4.00~K for the initial configuration (uniformly $d=0.593$) to 0.901~K. It should be noted that, in the variable-lattice-density optimization, intermediate design variable values in the range $0<d<1$ retain physical meaning because the design variable and the cylinder radius are in one-to-one correspondence through $r=(0.01+0.02d)$~m. The optimized design indicates that the design variable becomes smaller along the outer wall, whereas relatively large values appear around the center of the domain.

To examine sensitivity to the initial design, the baseline optimization was repeated using random distribution~\#1 in Fig.~\ref{validation} as the initial design, with all other settings unchanged. The final reduced-model objective value was 0.904~K, compared with 0.901~K for the uniform initial design. The optimized pin-fin radius distribution was similar to that obtained from the uniform initial design.

\begin{figure}
	\centering
	\includegraphics[width=.9\textwidth]{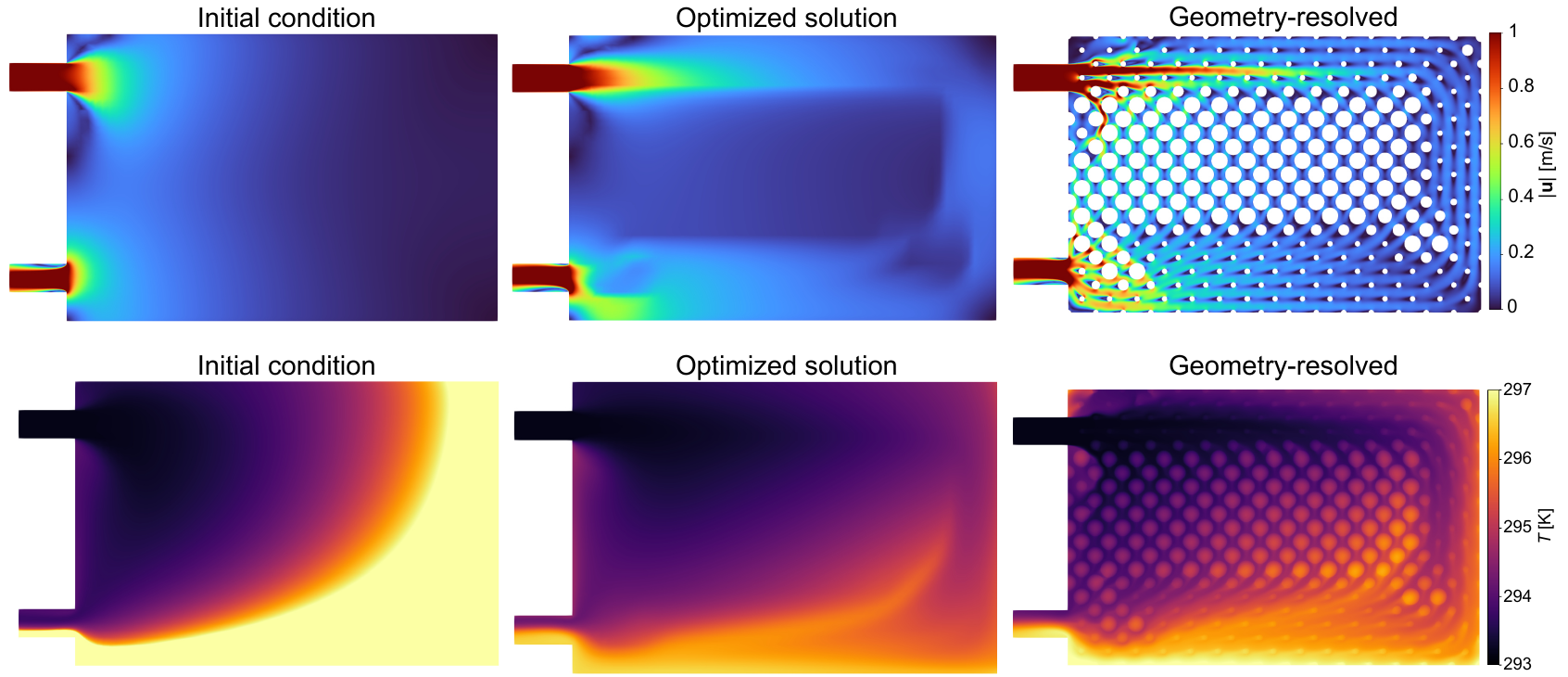}
	\caption{Velocity distributions (top) and temperature distributions (bottom) for the initial condition, optimized solution, and the geometry-resolved simulation results. The RM and GR panels show the Darcy speed and local fluid speed, respectively.}
	\label{initopt}
\end{figure}

The final design at the 100th optimization step was converted into an explicit cylinder array by dehomogenization and then analyzed using a RANS simulation with GR Mesh~4, containing 366,429 elements. The objective value decreased from 4.27~K to 0.942~K in the geometry-resolved simulation, while it decreased from 4.00~K to 0.901~K in the proposed method, indicating that the reconstructed design retains a substantial performance improvement. For the initial and optimized baseline designs, the wall-clock times required to obtain converged flow and temperature fields in a single analysis were 30~s and 36~s for the reduced model and 2667~s and 419~s for the geometry-resolved RANS model, respectively, corresponding to 98.9\% and 91.4\% reductions in computation time. The timing measurements were obtained on a workstation equipped with an Intel Core i9-14900K processor and 128 GB of memory. Figure~\protect\ref{initopt} shows the flow fields for the initial condition, the optimized solution, and the geometry-resolved RANS simulation of the reconstructed geometry. In the initial design, the velocity is low on the right side of the design domain, indicating that the flow does not adequately reach that region. As a result, the temperature is particularly high on the right side. In contrast, the flow in the optimized design shows high-velocity regions along the outer wall. This is achieved by the formation of low-resistance flow paths along the outer wall, where the design variable is small (i.e., the pin-fin radius is small). As a consequence, sufficient flow is supplied to the right side of the domain, resulting in a much more uniform temperature distribution over the whole domain. The geometry-resolved simulation of the reconstructed geometry also shows the formation of high-velocity regions along the outer wall, suggesting that the main flow-redistribution pattern obtained by the reduced model is preserved after dehomogenization.

\begin{figure}
	\centering
	\includegraphics[width=.9\textwidth]{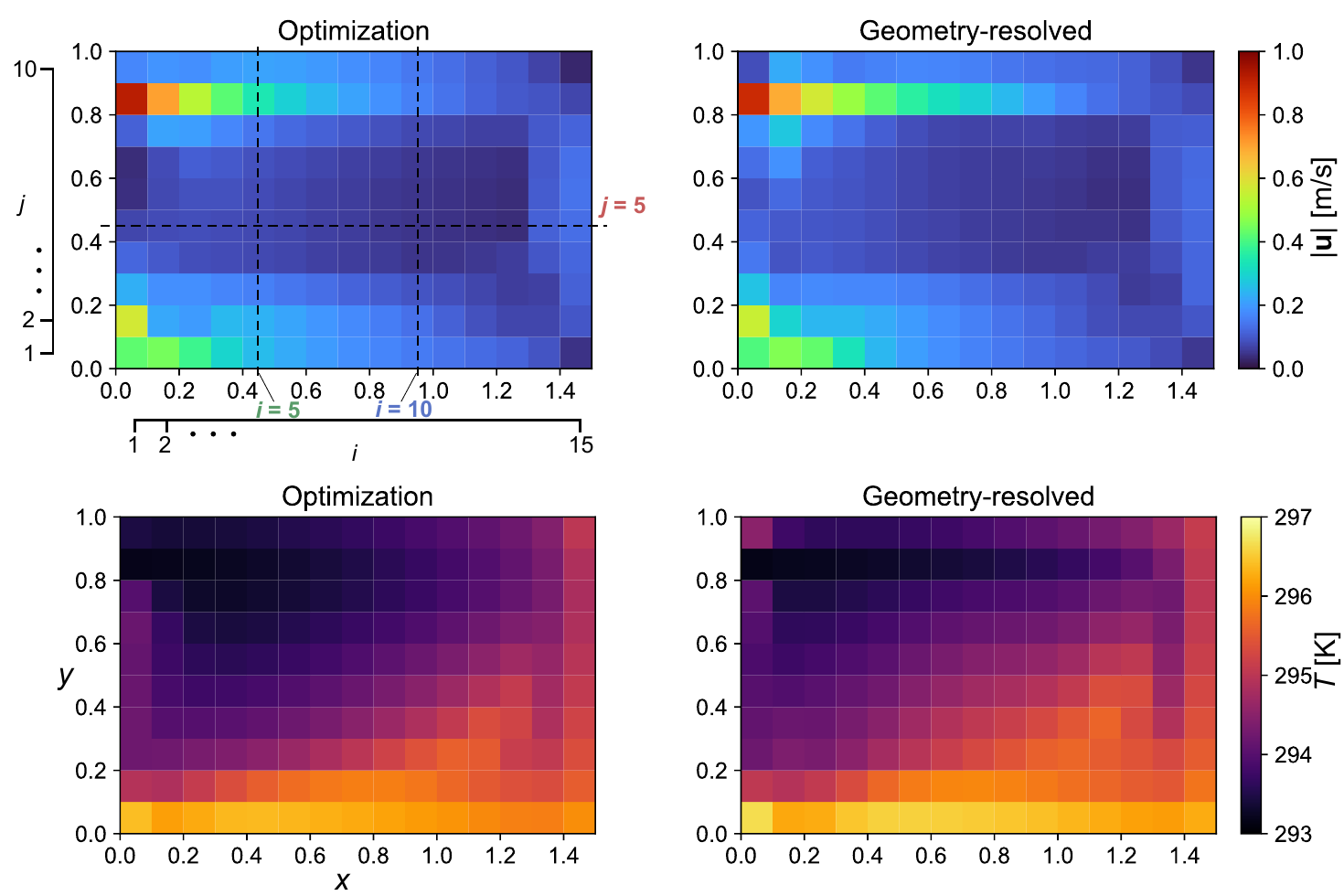}
	\caption{Comparisons of velocity distributions (top) and temperature distributions (bottom) between the optimization and geometry-resolved simulation.}
	\label{opthf}
\end{figure}

\begin{figure}
	\centering
	\includegraphics[width=.9\textwidth]{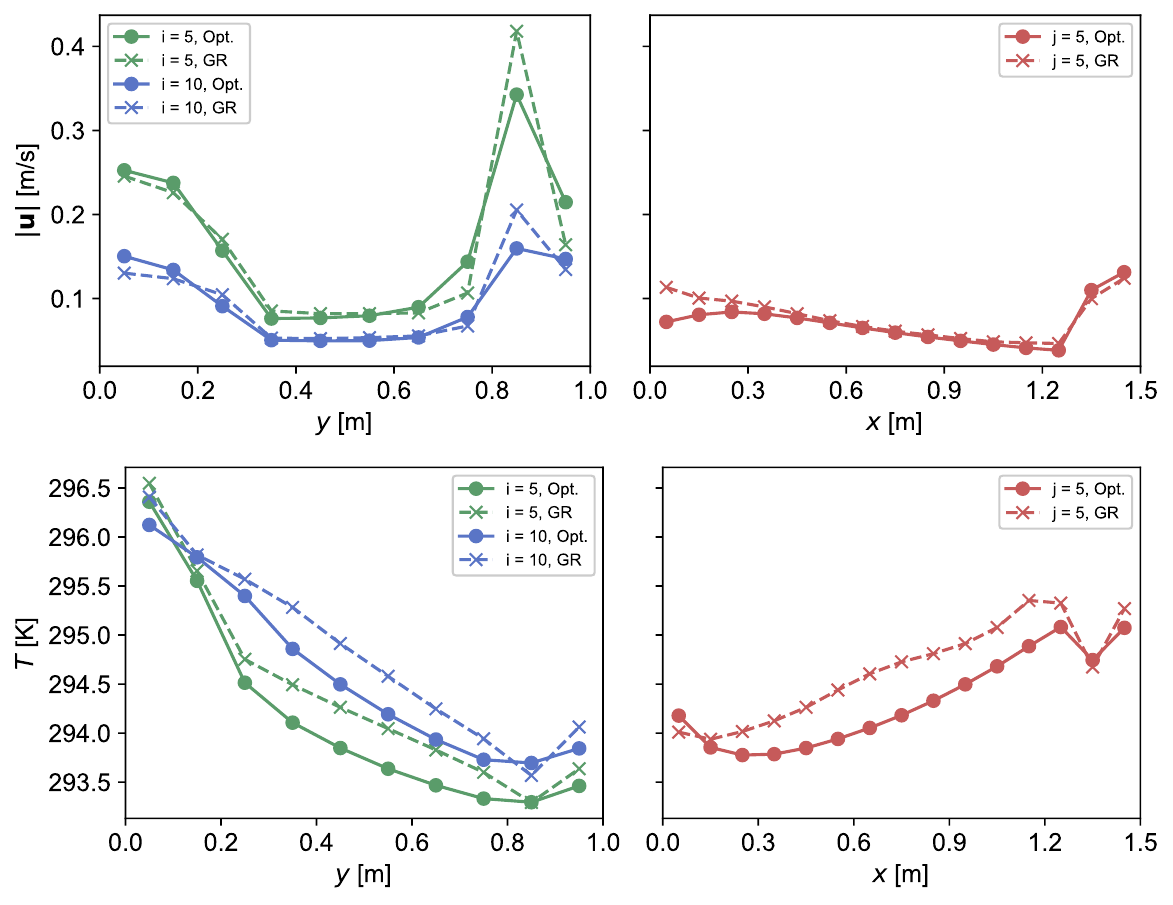}
	\caption{Comparisons of velocity and temperature profiles along $i = 5$, $10$, and $j = 5$ between the optimization (Opt.) and geometry-resolved simulation (GR).}
	\label{lineoptGR}
\end{figure}

Because the proposed method describes unit-cell-averaged macroscopic fields rather than cylinder-scale local variations, the geometry-resolved results were spatially averaged over a $15\times 10$ partition before comparison with the optimization results. Figure~\protect\ref{opthf} compares the macroscopic flow fields obtained from the optimization with the unit-cell-averaged geometry-resolved RANS simulation results. Here, $(i,j)$ denotes the index of the coarse mesh element counted from the lower-left element in the $x$- and $y$-directions. The averaged velocity distributions show good qualitative agreement between the two calculations, especially in the formation of high-velocity regions along the outer walls. The temperature distributions also show the same overall tendency toward a more uniform field. Figure~\ref{lineoptGR} provides a more detailed comparison of the velocity and temperature profiles along $i=5$, $i=10$, and $j=5$. The velocity profiles obtained by the proposed method agree well with those of the geometry-resolved computation. For the temperature field, although the pointwise agreement is weaker than that for the velocity field, the overall profile trends are captured. These results indicate that the proposed method captures the dominant macroscopic flow redistribution and reasonably predicts the associated thermal transport trend.

\subsection{Optimization with a pressure-drop constraint}
\label{sec:pressure_constraint}

An additional optimization was performed under the baseline condition $U_{\mathrm{in}}=1$~m/s with the pressure-drop constraint $\Delta p_{\mathrm{lim}}=1100~\mathrm{Pa}$. The objective function, porosity constraint, initial design, boundary conditions, heat source, meshes, and optimization settings were the same as those in Section~\ref{sec:baseline}. The obtained design was reconstructed into an explicit cylinder array and re-evaluated by geometry-resolved RANS analysis.

\begin{figure}
\centering
\includegraphics[width=.9\textwidth]{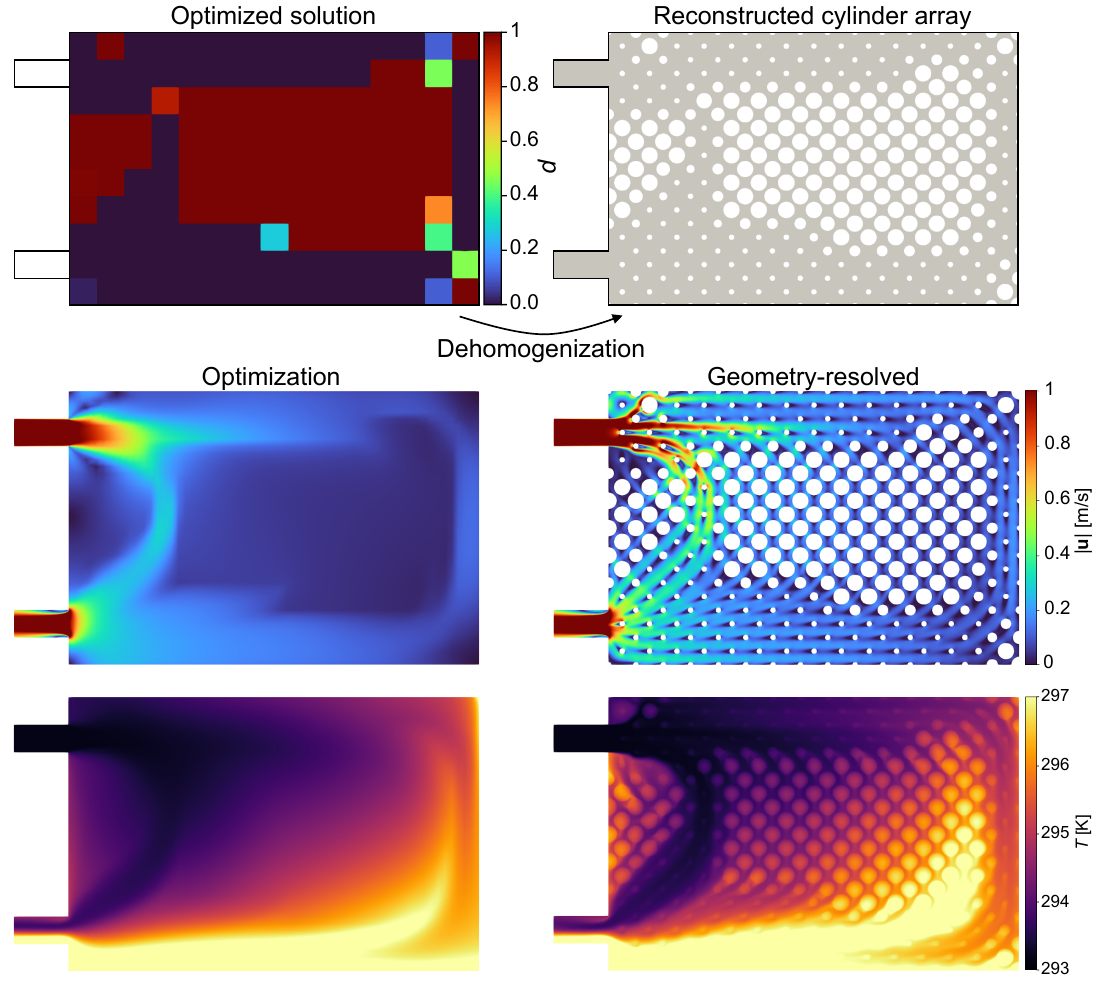}
\caption{Results obtained with a pressure-drop limit of 1100~Pa: optimized design variable distribution and reconstructed geometry (top), and velocity (middle) and temperature (bottom) distributions obtained from the reduced model (left) and geometry-resolved RANS analysis (right).}
\label{fig:pressure_constraint}
\end{figure}

Figure~\ref{fig:pressure_constraint} shows the optimized design variable distribution and the geometry-resolved velocity and temperature fields. Low-design-variable regions form a relatively short connection between the inlet and outlet, while a longer route extends along the outer walls. The coexistence of these routes is consistent with the need to limit pressure drop while distributing the flow over the design domain to improve temperature uniformity.

Table~\ref{tab:pressure_constraint} compares the objective values and pressure drops with those of the baseline design obtained without the pressure-drop constraint. Here, $J_{\mathrm{RM}}$ and $\Delta p_{\mathrm{RM}}$ are evaluated using the reduced model, whereas $J_{\mathrm{GR}}$ and $\Delta p_{\mathrm{GR}}$ are obtained from geometry-resolved RANS analysis of each reconstructed design.

\begin{table}
\centering
\caption{Objective values and pressure drops of the designs optimized with and without the pressure-drop constraint at $U_{\mathrm{in}}=1$~m/s.}
\label{tab:pressure_constraint}
\begin{tabular}{lrrrr}
\toprule
Design & $J_{\mathrm{RM}}$ [K] & $J_{\mathrm{GR}}$ [K] & $\Delta p_{\mathrm{RM}}$ [Pa] & $\Delta p_{\mathrm{GR}}$ [Pa] \\
\midrule
Without pressure-drop constraint & 0.901 & 0.942 & 1753 & 1213 \\
With pressure-drop constraint & 1.32 & 1.36 & 1100 & 898.4 \\
\bottomrule
\end{tabular}
\end{table}

The pressure-drop constraint was active in the reduced-model optimization, with a final pressure drop of 1100~Pa. For the reconstructed geometry, the geometry-resolved RANS analysis gave a pressure drop of 898.4~Pa, confirming that the prescribed limit was also satisfied. Although the objective values obtained by the two models were close, the reduced model overpredicted the pressure drop by 22.4\% relative to the geometry-resolved result. These results indicate that the proposed method can be applied to design exploration with an additional hydraulic constraint, with the resulting design assessed by geometry-resolved analysis.

\subsection{Comparison with simplified flow models}

To examine how the macroscopic flow and heat-transfer models affect the optimized design, additional optimizations were performed using two simplified flow models under the same optimization settings as the baseline case. In the Darcy--Forchheimer-only model, Eq.~\eqref{eq:df} was used directly as the macroscopic momentum equation instead of Eq.~\eqref{eq:ns}; therefore, the convective and diffusive momentum-transport terms included in Eq.~\eqref{eq:ns} were omitted. In the laminar flow model, Eqs.~\eqref{eq:ns},~\eqref{eq:energy}, and~\eqref{eq:keff} were used with $\mu_t=0$. In both simplified-model optimizations, the same effective properties $\kappa(d)$ and $\beta(d)$ identified from the unit-cell calculations were used, and the objective function, constraint, initial design, and operating conditions were kept identical to those in the baseline case. The optimized designs obtained by the simplified models were also reconstructed into explicit cylinder arrays by dehomogenization and re-evaluated by geometry-resolved RANS simulations.

\begin{figure}
    \centering
    \includegraphics[width=.9\textwidth]{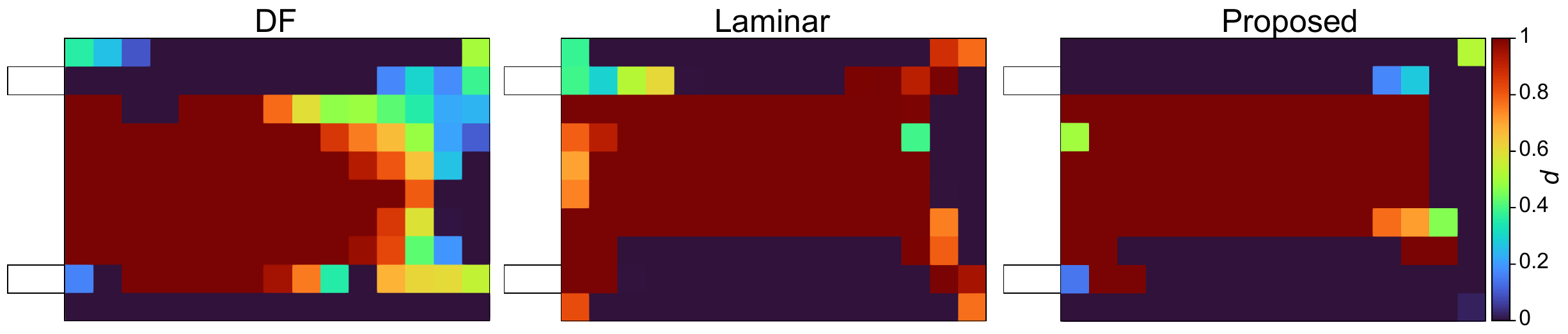}
    \caption{Comparison of optimized design variable distributions obtained using the Darcy--Forchheimer-only model (DF), laminar flow model (Laminar), and proposed model.}
    \label{dflamturb}
\end{figure}

Figure~\ref{dflamturb} compares the optimized design variable distributions obtained by the three models. The overall distributions are qualitatively similar: all models generate regions with small design variable values along the outer wall, forming low-resistance flow paths, while relatively large design variable values remain around the central part. This indicates that the main design tendency is captured even by the simplified models.

However, the objective values summarized in Table~\ref{tab:dflamturb_obj} show clear quantitative differences. The Darcy--Forchheimer-only model gives the largest geometry-resolved objective value among the three models. Although the laminar flow model and the proposed model yield very similar design variable distributions, the proposed model gives a lower objective value than the laminar flow model, both in the optimization model and in the geometry-resolved simulation. The absolute difference between the reduced-model and geometry-resolved objective values is 0.041~K for the proposed model, compared with 0.17~K for each simplified model. Thus, the proposed model gives closer agreement with the geometry-resolved evaluation in this comparison, even though the main design tendency is similar among the models.

\begin{table}
\centering
\caption{Comparison of objective values obtained using the Darcy--Forchheimer-only, laminar flow, and proposed models.}
\label{tab:dflamturb_obj}
\begin{tabular}{lcc}
\toprule
Method & \multicolumn{2}{c}{Final objective value, $J$ [K]} \\
\cmidrule(lr){2-3}
 & Optimization & Geometry-resolved RANS \\
\midrule
Darcy--Forchheimer-only model & 1.04 & 1.21 \\
Laminar flow model & 1.32 & 1.15 \\
Proposed model & 0.901 & 0.942 \\
\bottomrule
\end{tabular}
\end{table}

\subsection{Sensitivity to unit-cell size}

The sensitivity of the optimization result to the prescribed unit-cell size was also examined using the reduced model. The unit-cell size was reduced from $s=0.1$~m to $s=0.05$~m, so that the design domain was represented by $30\times20$ unit cells instead of $15\times10$ unit cells. The pin-fin radius was scaled geometrically as $r(d)=(0.005+0.01d)$~m, while the same operating conditions, objective function, and constraint were used. For this smaller unit cell, the effective permeability $\kappa(d)$ and the Forchheimer coefficient $\beta(d)$ were re-identified from unit-cell calculations, and the optimization was carried out in the same design domain as in the baseline case.

\begin{figure}
    \centering
    \includegraphics[width=.9\textwidth]{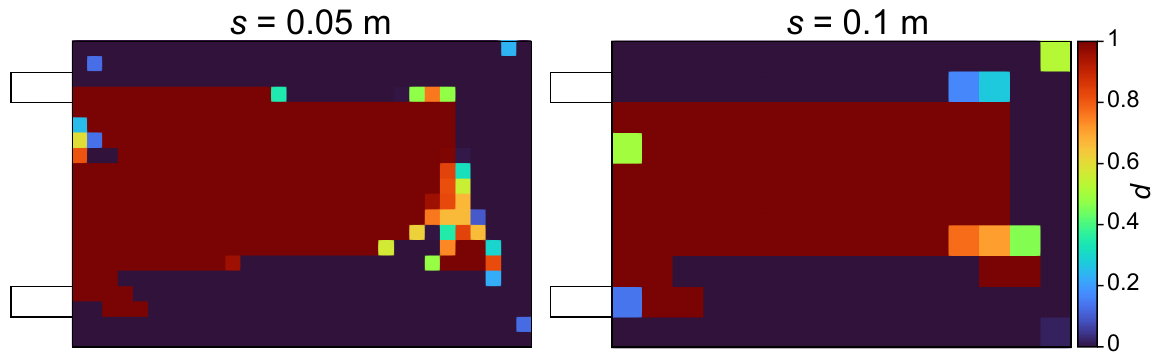}
    \caption{Comparison of optimized design variable distributions obtained with $s=0.05$ m and $s=0.1$ m.}
    \label{fig:smaller_unit_cell}
\end{figure}

Figure~\ref{fig:smaller_unit_cell} shows the optimized design variable distributions obtained with the smaller and baseline unit cells. The final objective value obtained with the smaller unit cell was 0.816~K, whereas that for the baseline unit cell was 0.901~K. Although local differences appeared at the cell scale owing to the increased number of unit cells, the overall trend of the optimized design variable distribution was similar to that in the baseline case, particularly in the formation of low-resistance regions near the outer walls. This result suggests that the overall tendency to form low-resistance regions near the outer walls is mainly governed by macroscopic flow redistribution, although the final objective value shows some quantitative dependence on the prescribed unit-cell size.

\subsection{Effects of inlet velocity}
\label{sec:inlet_velocity}

\begin{figure}
	\centering
	\includegraphics[width=1.\textwidth]{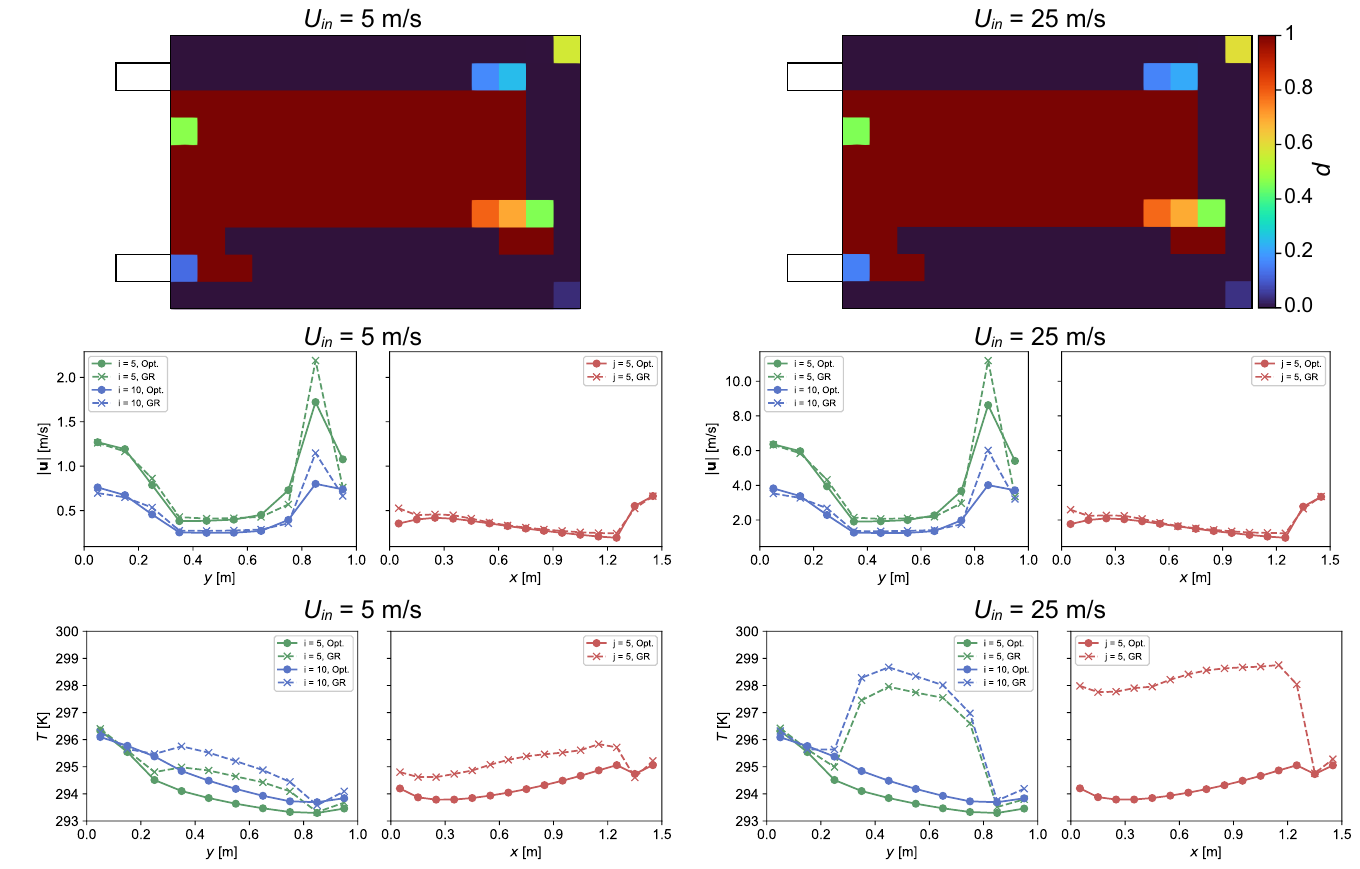}
	\caption{Optimized design variable distributions for $U_{\mathrm{in}} = 5$ and $25\ \mathrm{m/s}$, and comparison of the velocity and temperature profiles at $i = 5$, $i = 10$, and $j = 5$ between the optimization (Opt.) and geometry-resolved simulations (GR).}
	\label{uin525}
\end{figure}

Optimization was next performed under higher-Reynolds-number conditions with inlet velocities $U_{\mathrm{in}}=5$ and $25$~m/s, corresponding to maximum permeability-based Reynolds numbers $Re_{\kappa}=54,232$ and 272,784, respectively. The heat source $\dot{q}$ was increased in proportion to the inlet velocity, maintaining the same heat input per unit mass flow rate. This scaling keeps the reference temperature rise $\Delta T_{\mathrm{ref}}=\dot{Q}'/(\dot{m}'c_{p,f})$ constant, where $\dot{Q}'$ and $\dot{m}'$ are the total heat input and mass flow rate per unit depth, respectively. With constant properties and no thermal feedback on the flow, the energy equations are linear in the temperature rise for a fixed geometry and inlet velocity. Changing only the heat-source amplitude therefore scales the temperature deviations and $J$ by the same factor, without changing design rankings or the relative discrepancy between the reduced-model and geometry-resolved RANS objective values. The following comparisons examine the inlet-velocity dependence under this fixed heat-input-to-heat-capacity-rate ratio. The local Darcy velocities remain below the maximum Darcy velocity used to identify $\beta$ for the corresponding cylinder diameter over most of the optimized design domain at 5 m/s (97.1\%), whereas this fraction falls to 39.3\% at 25 m/s. The latter case therefore examines whether the method can still identify an improved design beyond the identification range. The optimized designs were then evaluated using geometry-resolved RANS. Figure~\protect\ref{uin525} shows the design variable distributions at the 100th optimization step for each condition, together with comparisons of velocity and temperature distributions between the optimization and the geometry-resolved simulations. The design variable distributions are similar for all conditions, and the overall tendencies of the velocity and temperature fields obtained by the optimization are also similar. However, the agreement between the predicted temperature fields deteriorates as the inlet velocity increases, although the trends of the velocity distributions remain broadly consistent between the optimization and the geometry-resolved simulations.

\begin{figure}
	\centering
	\includegraphics[width=1.\textwidth]{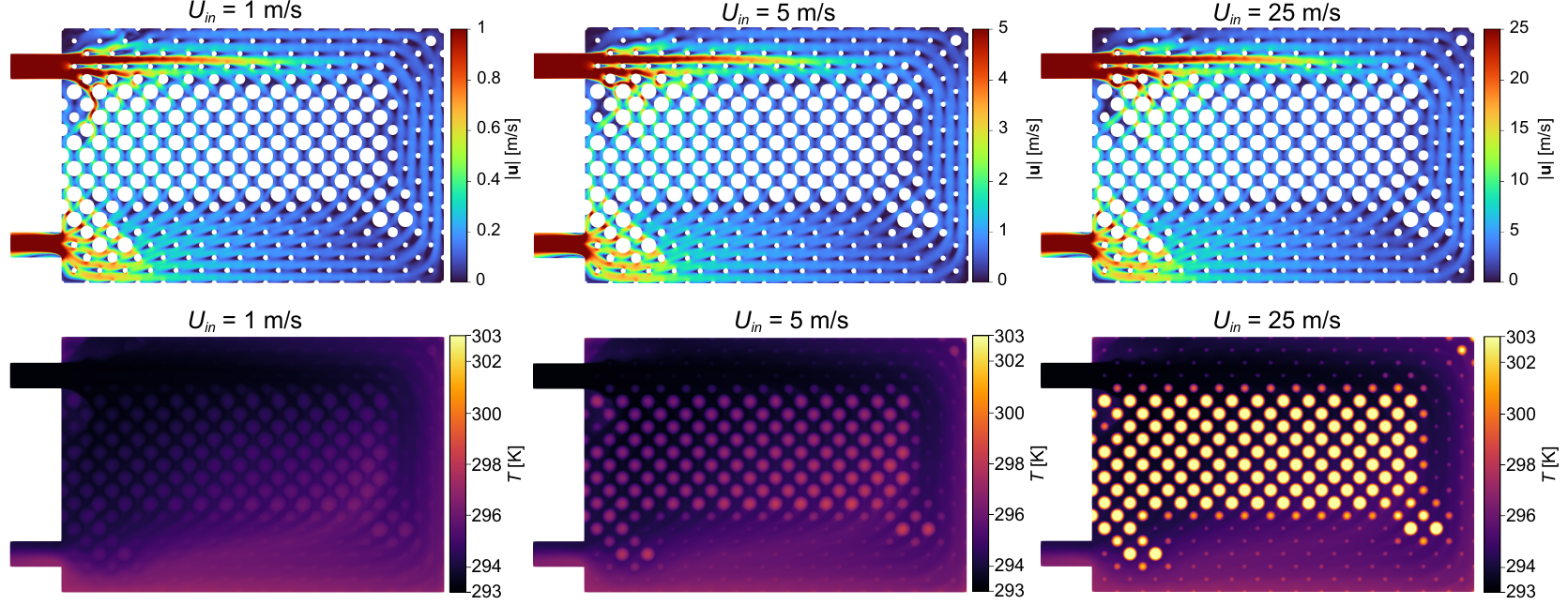}
	\caption{Comparison of velocity and temperature distributions obtained from the geometry-resolved simulations for $U_{\mathrm{in}} = 1$, 5, and $25\ \mathrm{m/s}$.}
	\label{uin125}
\end{figure}

Figure~\protect\ref{uin125} compares the velocity and temperature fields for $U_{\mathrm{in}}=1$, 5, and 25~m/s. The velocity distributions show a similar tendency for all inlet velocities: the main flow paths are formed near the outer walls, whereas the central region remains relatively low in velocity. In contrast, the temperature distribution becomes increasingly nonuniform as the inlet velocity and heat-generation rate increase. In particular, the geometry-resolved simulations show pronounced temperature peaks inside the cylinders at higher inlet velocities, especially for $U_{\mathrm{in}}=25$~m/s. Nevertheless, compared with the initial design, the objective value at $U_{\mathrm{in}}=5$~m/s was reduced from 3.85~K to 1.07~K in the geometry-resolved simulation and from 3.87~K to 0.893~K in the optimization. At $U_{\mathrm{in}}=25$~m/s, it decreased from 4.29~K to 3.49~K in the geometry-resolved simulation and from 3.83~K to 0.889~K in the optimization. The reduction confirmed by the geometry-resolved simulation was therefore 18.6\%, substantially smaller than the 76.8\% predicted by the reduced model. The good agreement in the macroscopic velocity distributions indicates that the reduced model captures the dominant flow redistribution even at 25~m/s. The much smaller reduction in $J$ found in the geometry-resolved analysis appears to arise mainly from the high temperatures within the cylinders, as shown in Fig.~\ref{uin125}. To examine the origin of this discrepancy, a post-processing decomposition of the geometry-resolved temperature field was performed. Within each pin, the component $T_{\mathrm{cond}}(\mathbf{x})=\dot q\left[r_p^2-\|\mathbf{x}-\mathbf{x}_p\|^2\right]/(4k_s)$ satisfies the steady conduction equation with uniform internal heating and vanishes at the pin surface, where $r_p$ and $\mathbf{x}_p$ denote the radius and center position of pin $p$, respectively, and $\|\mathbf{x}-\mathbf{x}_p\|$ is the distance from the evaluation point $\mathbf{x}$ to the pin center. This component was set to zero in the fluid. Evaluating the objective for $T_{\mathrm{GR}}-T_{\mathrm{cond}}$ gave 0.918~K, close to the reduced-model value of 0.889~K. This result supports the temperature rise caused by internal heating and finite thermal conduction within the pins as the main contributor to the discrepancy at $U_{\mathrm{in}}=25$~m/s. These local temperature variations are not resolved by the single-temperature model, which consequently overpredicts the improvement in temperature uniformity. A two-temperature model could better represent these solid--fluid temperature differences. We retain the single-temperature formulation for efficient candidate-design exploration, with the resulting designs assessed by geometry-resolved analyses. Extension to a two-temperature model, including identification of the interfacial heat-transfer coefficient, is left for future work.

\subsection{Effects of outlet position}

\begin{figure}
	\centering
	\includegraphics[width=.9\textwidth]{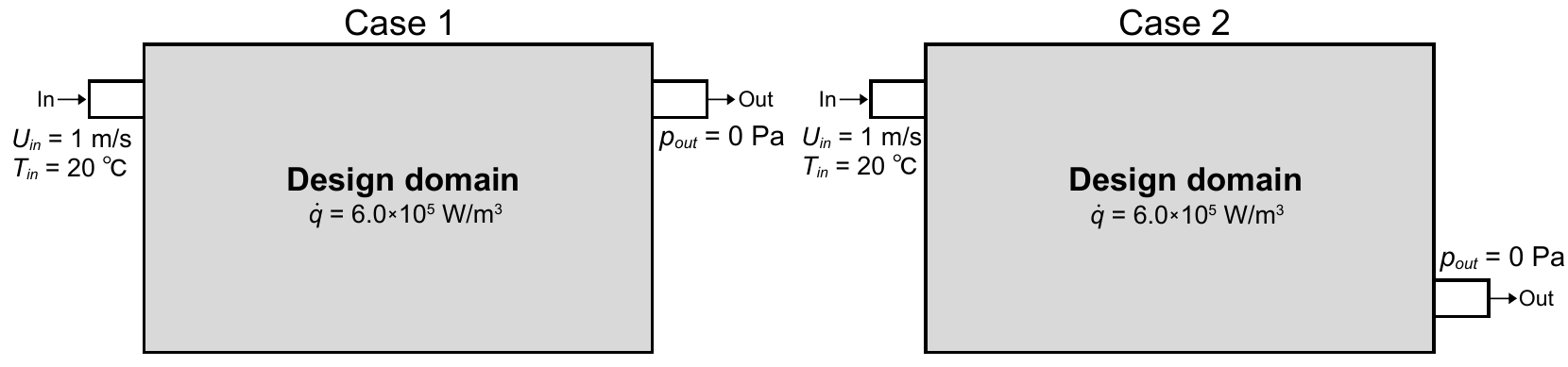}
	\caption{Two outlet configurations with different outlet positions (Case 1 and Case 2).}
	\label{compdomain2}
\end{figure}

This section examines whether the proposed method can still identify candidate designs when the global flow structure changes because of a change in outlet position. Two outlet arrangements were considered, as shown in Fig.~\protect\ref{compdomain2}. In Case~1, the outlet was placed on the side opposite the inlet and aligned with it. In Case~2, the outlet was placed on the lower part of the right boundary, opposite to the outlet location in the baseline geometry. For each case, the same optimization conditions as those in Section~3.2 were applied. The obtained designs were then reconstructed into explicit cylinder arrays by dehomogenization and re-evaluated by geometry-resolved RANS analysis.

\begin{figure}
	\centering
	\includegraphics[width=1.\textwidth]{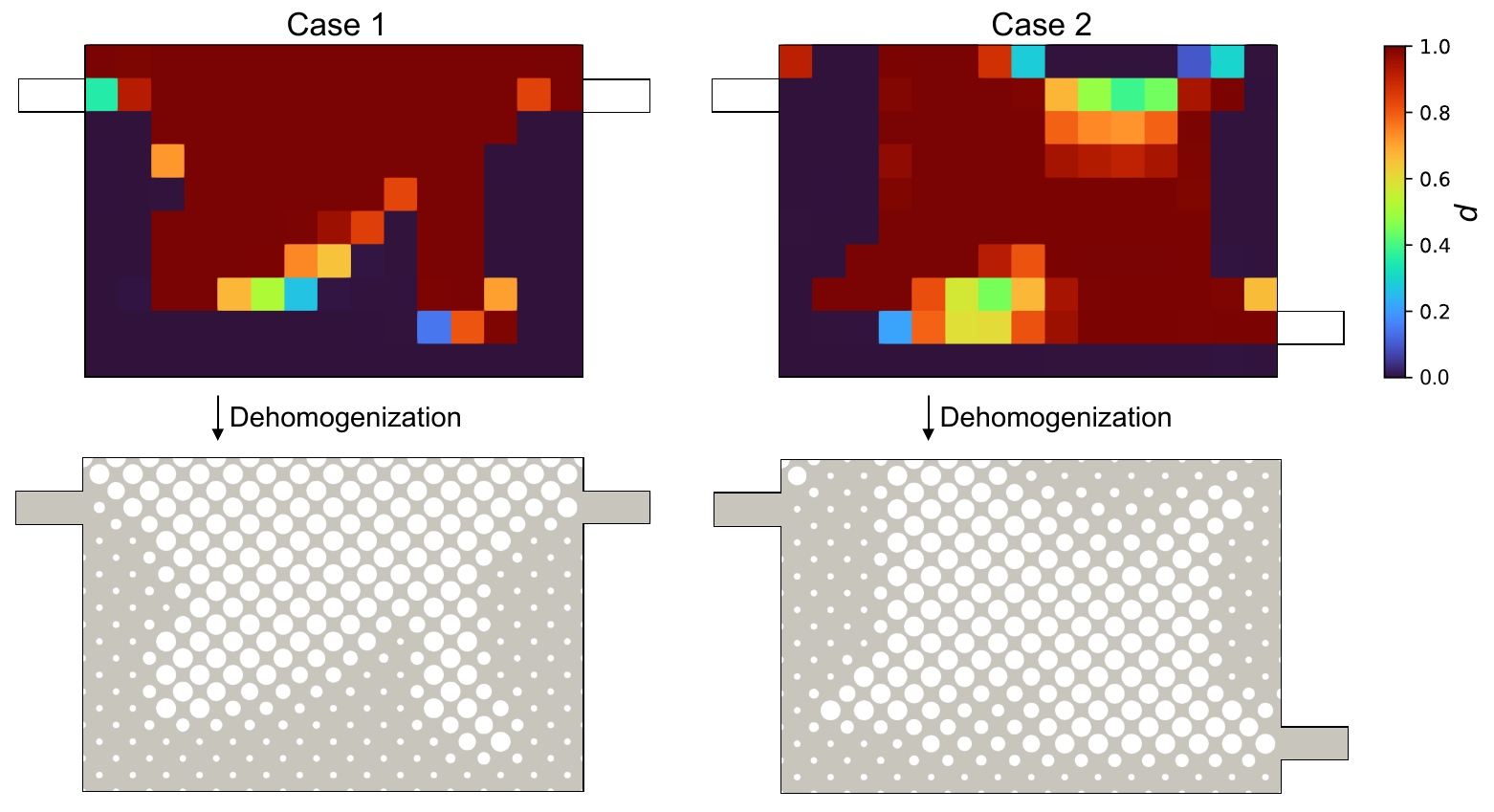}
	\caption{Optimized design variable distributions and the reconstructed cylinder arrays for Cases 1 and 2.}
	\label{outlet_d}
\end{figure}

Figure~\protect\ref{outlet_d} shows the optimized design variable distributions and the reconstructed cylinder arrays for Cases~1 and 2. In Case 1, a region with low design variable values appears near the inlet and extends downward along the wall toward the outlet. In Case 2, a similar low-design-variable region extends downward along the wall from the inlet, while another such region also appears in the upper-right part of the domain. In the geometry-resolved RANS analysis of the reconstructed geometries, the objective value decreased from 1.92~K to 0.615~K for Case~1, and from 1.14~K to 0.701~K for Case 2. In the optimization based on the reduced model, the objective value decreased from 1.62~K to 0.615~K for Case~1 and from 0.997~K to 0.583~K for Case~2. These results indicate that the proposed method remains useful for exploring candidate designs under modified outlet configurations, although the quantitative accuracy of the objective value is limited.

\begin{figure}
	\centering
	\includegraphics[width=1.\textwidth]{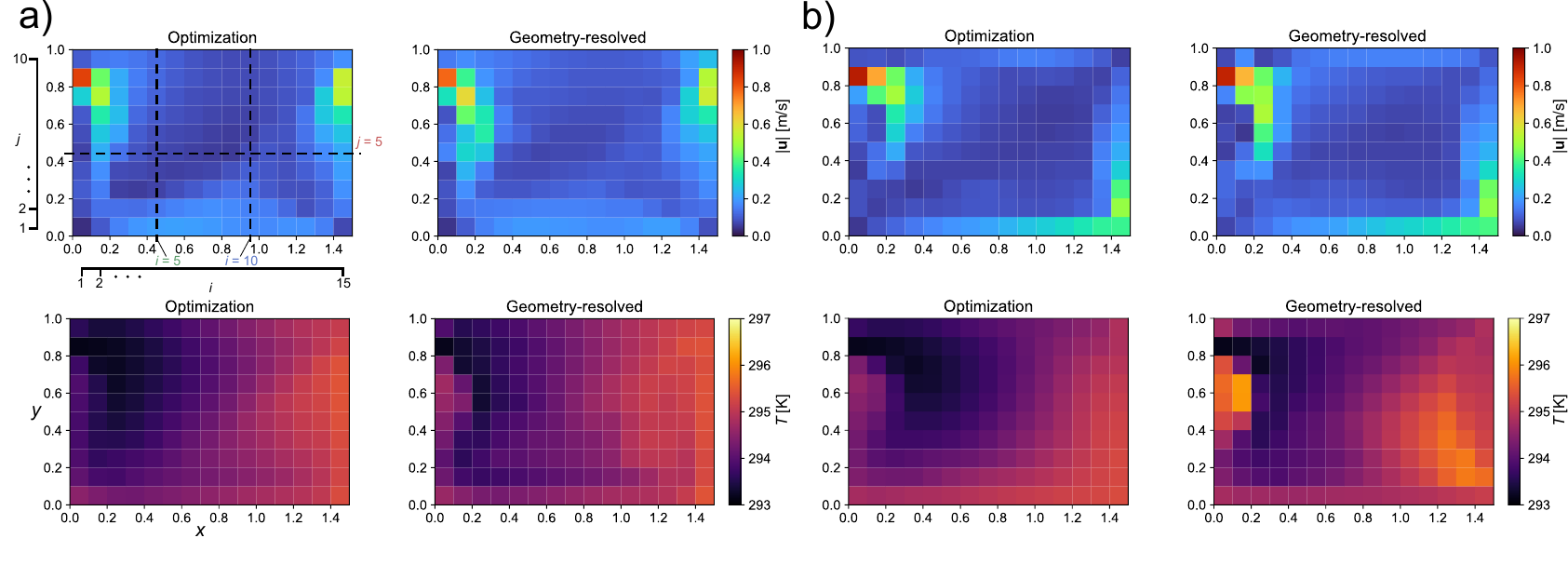}
	\caption{Comparison of velocity (top) and temperature (bottom) distributions between the optimization and geometry-resolved simulation after area-averaging over each unit cell: (a) Case 1 and (b) Case 2.}
	\label{avgff12}
\end{figure}

Figure~\protect\ref{avgff12} compares the velocity and temperature distributions obtained from the optimization and geometry-resolved simulations after area-averaging over each unit cell. In the velocity fields of both cases, the optimization results show similar overall trends to those of the geometry-resolved simulations. In contrast, the temperature distributions indicate that high-temperature regions appear near the inlet in the geometry-resolved simulations of both cases.

\begin{figure}
	\centering
	\includegraphics[width=1.\textwidth]{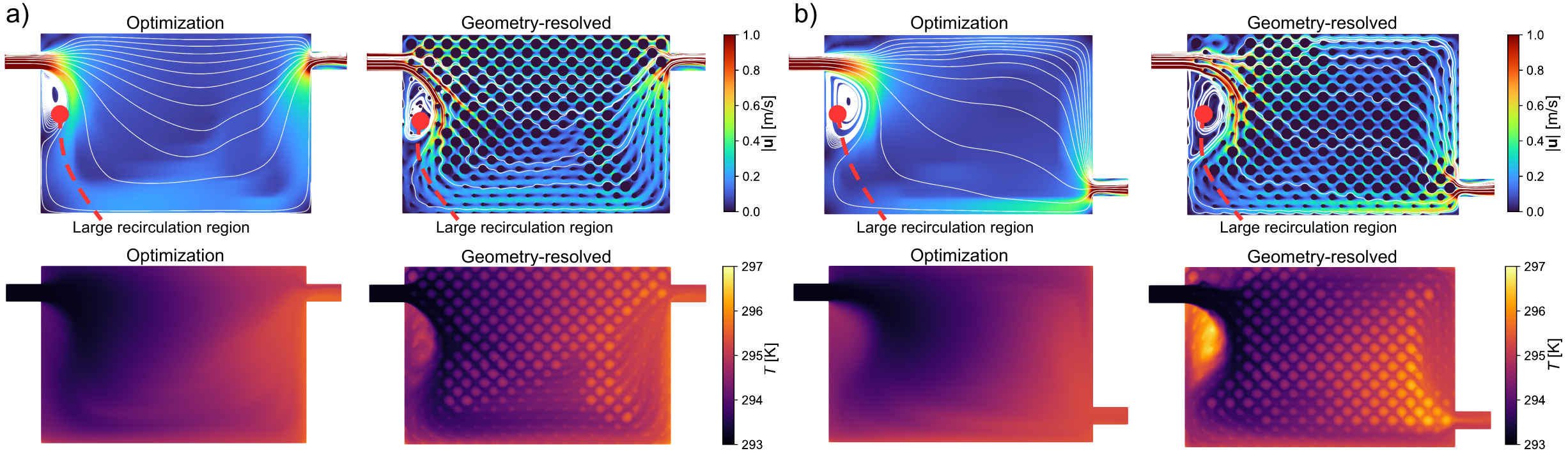}
	\caption{Comparison of velocity (top) and temperature (bottom) distributions between the optimization and geometry-resolved simulations: (a) Case 1 and (b) Case 2.}
	\label{veltemp12}
\end{figure}

Figure~\protect\ref{veltemp12} compares the flow fields obtained by the optimization and geometry-resolved simulations. In both analyses, the inflow separates from the wall of the inlet pipe, and large-diameter cylinders near the inlet block the incoming flow and deflect it downward, forming a large recirculation region. The temperature distribution shows that the reduced model underestimates the temperature peaks in that region. Because the local velocity decreases in the center of such a recirculation region, convective heat transport is weakened and a high-temperature region is more likely to form.

\begin{figure}
	\centering
	\includegraphics[width=.9\textwidth]{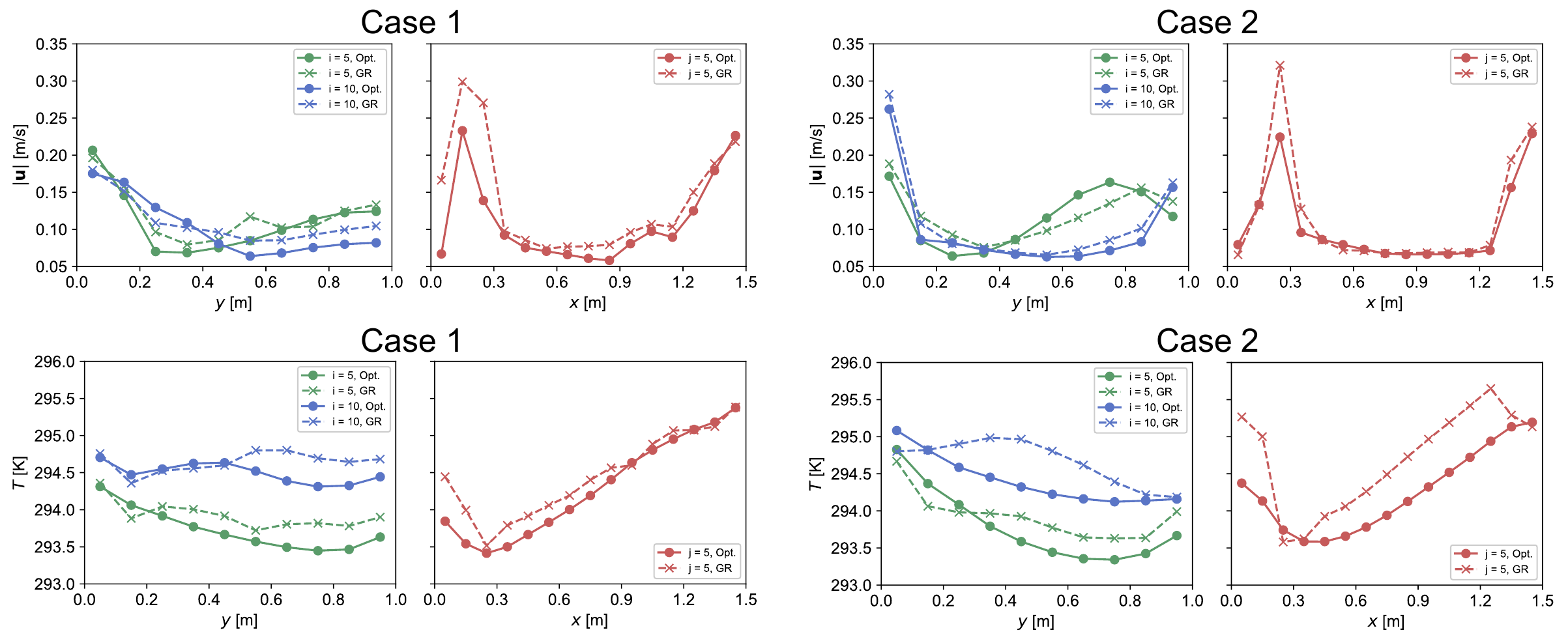}
	\caption{Comparison of velocity profiles (top) and temperature profiles (bottom) along $i = 5$, $10$, and $j = 5$ for Cases 1 and 2 between the optimization (Opt.) and geometry-resolved simulations (GR).}
	\label{lineveltemp12}
\end{figure}

Figure~\protect\ref{lineveltemp12} compares the velocity and temperature distributions along $i=5$, $i=10$, and $j=5$ for Cases~1 and 2. Overall, the distributions obtained by the optimization are broadly consistent with those obtained by the geometry-resolved simulation, and the main trends of the velocity and temperature fields are reasonably captured. However, relatively large discrepancies are observed near the recirculation region on the inlet side, particularly around $(i,j)=(1\text{--}4,5)$. These discrepancies indicate that the homogenization-based reduced model does not fully reproduce the inlet-side separation, recirculation, and associated temperature peaks predicted by geometry-resolved RANS. The reduced model describes macroscopic flow and heat transfer without explicitly resolving individual pin-fin geometries, whereas geometry-resolved RANS explicitly accounts for these geometries when computing the mean flow and temperature fields. In addition, the inlet and outlet widths are comparable to the unit-cell size, so the scale separation required by the homogenized description is not satisfied in their immediate vicinity, limiting the accuracy of the local predictions.

These results indicate that, although the homogenization-based reduced model cannot quantitatively resolve local separation and recirculation in detail, it can still identify global flow patterns that are beneficial for temperature uniformity. Therefore, the optimized geometries can still be regarded as reasonable candidate designs.

Table~\ref{tab:gr_performance} summarizes the geometry-resolved values of $J$, $\Delta p$, and $T_{\max}$ for the designs used in the performance comparisons. Here, $T_{\max}$ is the maximum temperature in $\Omega$, evaluated from the geometry-resolved temperature field before unit-cell averaging. At each operating condition, the inlet flow rate, inlet temperature, and total heat input are identical for the designs compared.

\begin{table}[pos=htbp]
\centering
\caption{Geometry-resolved temperature nonuniformity, pressure drop, and maximum temperature of the designs used in the performance comparisons.}
\label{tab:gr_performance}
\small
\setlength{\tabcolsep}{4pt}
\begin{tabular*}{\linewidth}{@{\extracolsep{\fill}}lrrr@{}}
\toprule
Design & $J_{\mathrm{GR}}$ & $\Delta p_{\mathrm{GR}}$ & $T_{\max}$ \\
       & [K] & [Pa] & [K] \\
\midrule
\multicolumn{4}{@{}l}{Baseline geometry, $U_{\mathrm{in}}=1$~m/s} \\
Initial & 4.27 & $1.407\times10^3$ & 313.2 \\
Optimized (proposed model) & 0.942 & $1.213\times10^3$ & 297.2 \\
Optimized (pressure constrained) & 1.36 & $8.984\times10^2$ & 298.7 \\
Optimized (Darcy--Forchheimer-only) & 1.21 & $1.211\times10^3$ & 299.3 \\
Optimized (laminar model) & 1.15 & $1.690\times10^3$ & 298.1 \\
\midrule
\multicolumn{4}{@{}l}{Baseline geometry, $U_{\mathrm{in}}=5$~m/s} \\
Initial & 3.85 & $3.504\times10^4$ & 312.6 \\
Optimized (proposed model) & 1.07 & $2.928\times10^4$ & 298.4 \\
\midrule
\multicolumn{4}{@{}l}{Baseline geometry, $U_{\mathrm{in}}=25$~m/s} \\
Initial & 4.29 & $8.687\times10^5$ & 318.0 \\
Optimized (proposed model) & 3.49 & $7.264\times10^5$ & 309.8 \\
\midrule
\multicolumn{4}{@{}l}{Outlet Case~1, $U_{\mathrm{in}}=1$~m/s} \\
Initial & 1.92 & $1.551\times10^3$ & 301.8 \\
Optimized (proposed model) & 0.615 & $2.225\times10^3$ & 296.1 \\
\midrule
\multicolumn{4}{@{}l}{Outlet Case~2, $U_{\mathrm{in}}=1$~m/s} \\
Initial & 1.14 & $1.537\times10^3$ & 299.2 \\
Optimized (proposed model) & 0.701 & $1.627\times10^3$ & 296.8 \\
\bottomrule
\end{tabular*}
\end{table}

For all five operating configurations, the designs optimized using the proposed model reduced both $J$ and $T_{\max}$ relative to the corresponding initial designs. The pressure drop also decreased for the baseline geometry at all three inlet velocities, whereas it increased by 43.5\% and 5.9\% for outlet Cases~1 and~2, respectively. These results demonstrate that improved temperature uniformity does not necessarily imply reduced hydraulic cost.

\section{Conclusions}
In this study, variable-lattice-density optimization was extended to high-Reynolds-number pin-fin heat-sink design by combining unit-cell identification of Darcy--Forchheimer coefficients, a turbulent porous-flow model, and a dual-mesh optimization framework. The proposed framework enables the pin-fin radius distribution to be optimized on a unit-cell-scale design mesh while solving the macroscopic flow and temperature fields on a finer computational mesh.

The baseline optimization confirmed that the temperature-uniformity improvement obtained with the reduced model was retained after reconstruction: the objective value decreased from 4.27 K to 0.942 K in the geometry-resolved analysis and from 4.00 K to 0.901 K in the reduced model. The optimized design formed low-resistance flow paths along the outer wall, redistributed the flow over the domain, and improved temperature uniformity, with good agreement in the global velocity and temperature trends. Comparisons with simplified flow models and a smaller unit cell showed the same overall design tendency, indicating that it is governed by macroscopic flow redistribution.

An additional optimization of temperature uniformity with a pressure-drop constraint yielded a reconstructed design that satisfied the prescribed limit. This result highlights the importance of distributing flow between the shorter inlet-to-outlet connection and the peripheral routes to satisfy the pressure-drop limit while improving temperature uniformity relative to the initial design.

Agreement in the thermal objective deteriorated at higher inlet velocities, while modified outlet positions exposed local discrepancies near separation and recirculation regions. At \(U_{\mathrm{in}} = 25\,\mathrm{m/s}\), the reduced model predicted an optimized objective value of 0.889 K, compared with 3.49 K obtained from geometry-resolved RANS, substantially underestimating temperature nonuniformity. Accordingly, the validated range for quantitative prediction of the thermal objective is limited to the tested conditions with \(U_{\mathrm{in}} \leq 5\,\mathrm{m/s}\). The reconstructed designs nevertheless retained improved temperature uniformity, including in the 25~m/s case beyond the resistance-identification range. Across the tested two-dimensional cases, these findings support the framework as a practical strategy for identifying promising pin-fin layouts through macroscopic design exploration followed by geometry-resolved RANS assessment.

\appendix
\section{Comparison of isotropic and direction-dependent inertial resistance models}
\label{app:directional_drag}

This appendix examines the influence of the inertial resistance model on design exploration. Two resistance models were compared: the isotropic model with a scalar Forchheimer coefficient identified from the unit-cell calculations at $0^\circ$, and a direction-dependent model constructed from the calculations at $0^\circ$, $20^\circ$, and $40^\circ$. The same permeability $\kappa(d)$ was used in both models. The comparison considers the relative ranking of prescribed designs and the objective values of the optimized designs evaluated using geometry-resolved RANS simulations.

\subsection{Identification and implementation of the direction-dependent model}
\label{app:directional_identification}

The inertial resistance in the isotropic model is
\begin{equation}
\mathbf{R}_{\mathrm{iso}}(d,\mathbf{u})
=\rho_f\beta(d)|\mathbf{u}|\mathbf{u},
\label{eq:app_isotropic_resistance}
\end{equation}
where $\beta(d)$ is identified as described in Section~\ref{sec:effective_properties}. For the direction-dependent model, the inertial resistance is evaluated as
\begin{equation}
\mathbf{R}_{F}(d,\mathbf{u})
=\rho_f|\mathbf{u}|\mathbf{H}(d,u_x,u_y),
\qquad
\mathbf{H}=\begin{pmatrix}H_x\\H_y\end{pmatrix},
\label{eq:directional_resistance}
\end{equation}
where $\mathbf{H}$ is constructed from the unit-cell resistance data below. The conventional scalar Forchheimer term is recovered for $\mathbf{H}=\beta\mathbf{u}$. In the present formulation, $\mathbf{R}_{F}$ need not be parallel to $\mathbf{u}$, and the dependence of the resistance on velocity magnitude is not restricted to a quadratic law over the entire data range.

Unit-cell RANS simulations were performed at driving orientations of $0^\circ$, $20^\circ$, and $40^\circ$ relative to the $x$-axis, using the same pressure differences specified in Section~\ref{sec:effective_properties} for each orientation. The imposed pressure differences across the two pairs of periodic boundaries were $\Delta p_n\cos\alpha_n$ and $\Delta p_n\sin\alpha_n$, respectively. For each unit-cell result $n$, the Darcy velocity was evaluated using Eq.~\eqref{eq:cell_darcy_velocity}. The linear Darcy contribution was subtracted from the imposed pressure-drop gradient to obtain the nonlinear resistance:
\begin{equation}
\mathbf{G}_n
=\frac{\Delta p_n}{s}
\begin{pmatrix}
\cos\alpha_n\\
\sin\alpha_n
\end{pmatrix},
\qquad
\mathbf{R}_{F,n}
=\mathbf{G}_n-\frac{\mu}{\kappa(d_n)}\mathbf{u}_n,
\label{eq:cell_resistance}
\end{equation}
where $\mathbf{G}_n$ is the driving pressure-gradient vector, $\alpha_n$ is the driving orientation, and $d_n$ is the design variable. The computed velocity components were used directly, allowing the velocity direction to differ from the driving pressure-gradient direction.

Both components of the nonlinear resistance were retained through the auxiliary vector
\begin{equation}
\mathbf{H}_n
=\frac{\mathbf{R}_{F,n}}{\rho_f|\mathbf{u}_n|},
\qquad |\mathbf{u}_n|>0.
\label{eq:resistance_data}
\end{equation}
The data were extended using the rotation and reflection symmetries of the square lattice:
\begin{equation}
\mathbf{u}'_n=\mathbf{Q}\mathbf{u}_n,
\qquad
\mathbf{H}'_n=\mathbf{Q}\mathbf{H}_n,
\label{eq:resistance_symmetry}
\end{equation}
where $\mathbf{Q}$ denotes a lattice symmetry transformation. The resulting values of $H_x$ and $H_y$ were interpolated piecewise linearly as functions of $(d,u_x,u_y)$, with $\mathbf{H}(d,0,0)=\mathbf{0}$. The local values of these functions were then used in Eq.~\eqref{eq:directional_resistance} to evaluate the inertial resistance in the macroscopic momentum equation. This procedure retains the magnitude and direction of the resistance obtained from the unit-cell calculations.
For the direction-dependent model, the scalar Forchheimer coefficient entering $P_{k,\mathrm{pm}}$ and $P_{\varepsilon,\mathrm{pm}}$ is evaluated from the projection of $\mathbf{H}$ onto the Darcy velocity:
\begin{equation}
\beta_{\mathrm{turb}}
=\frac{\mathbf{H}(d,u_x,u_y)\cdot\mathbf{u}}
{|\mathbf{u}|^2}.
\label{eq:turbulence_coefficient}
\end{equation}
This scalar coefficient is used in the porous-medium turbulence production terms, while the full resistance vector in Eq.~\eqref{eq:directional_resistance} is used in the momentum equation. Equation~\eqref{eq:directional_resistance} is written without the regularization of the velocity magnitude near zero velocity.
In the direction-dependent calculation, the scalar Forchheimer term $-\rho_f\beta|\mathbf{u}|\mathbf{u}$ in Eq.~\eqref{eq:ns} was replaced by $-\mathbf{R}_{F}(d,\mathbf{u})$, while the Darcy term was retained. This comparison includes differences in the velocity-magnitude dependence, interpolation, and treatment of the porous-medium turbulence production terms as well as the directional dependence of the inertial resistance.

\subsection{Comparison of design rankings and optimized designs}

The same thirty design variable distributions used in Section~\ref{sec:validation} were evaluated with both resistance models under the baseline condition $U_{\mathrm{in}}=1$~m/s. Both models yielded the same Spearman rank correlation coefficient of $r_s=0.685$ relative to the geometry-resolved RANS results. The identical coefficients indicate that the direction-dependent model did not improve the ranking of candidate designs in this comparison.

Optimizations were also carried out with both resistance models at $U_{\mathrm{in}}=1$ and 5~m/s. At each inlet velocity, the initial design, boundary conditions, heat source, porosity constraint, meshes, and optimization settings were the same for the two models. The heat source followed the settings in Sections~\ref{sec:baseline} and~\ref{sec:inlet_velocity}. The resulting designs were reconstructed into explicit cylinder arrays and re-evaluated using the same geometry-resolved RANS procedure and objective-function definition. Table~\ref{tab:directional_drag} summarizes the objective values and pressure drops, where $J_{\mathrm{RM}}$ is evaluated using the reduced model employed in each optimization and $J_{\mathrm{GR}}$ is evaluated using geometry-resolved RANS.

\begin{table}
\centering
\caption{Objective values and pressure drops of the designs optimized using the isotropic and direction-dependent resistance models.}
\label{tab:directional_drag}
\begin{tabular}{clcccc}
\toprule
$U_{\mathrm{in}}$ [m/s] & Resistance model & $J_{\mathrm{RM}}$ [K] & $J_{\mathrm{GR}}$ [K] & $\Delta p_{\mathrm{RM}}$ [Pa] & $\Delta p_{\mathrm{GR}}$ [Pa]  \\
\midrule
1 & Isotropic & 0.901 & 0.942 & $1.753\times10^3$ & $1.213\times10^3$ \\
1 & Direction-dependent & 0.846 & 0.735 & $2.441\times10^3$ & $2.112\times10^3$ \\
5 & Isotropic & 0.893 & 1.07 & $4.330\times10^4$ & $2.928\times10^4$ \\
5 & Direction-dependent & 0.598 & 1.30 & $3.855\times10^4$ & $4.298\times10^4$ \\
\bottomrule
\end{tabular}
\end{table}

At $U_{\mathrm{in}}=1$~m/s, the design obtained with the direction-dependent model gave a lower geometry-resolved objective value than that obtained with the isotropic model, although its pressure drop was 74\% higher (2112 versus 1213 Pa). At $U_{\mathrm{in}}=5$~m/s, the direction-dependent design gave a higher $J_{\mathrm{GR}}$ than the isotropic model. The direction-dependent model also yielded larger discrepancies between the reduced-model and geometry-resolved RANS objective values than the isotropic model at both inlet velocities. Thus, incorporating the present direction-dependent resistance model can affect the optimized design, but it does not systematically improve the prediction of $J$ or reduce $J_{\mathrm{GR}}$ for the optimized designs over the tested conditions. We therefore retain the isotropic model in the main study as a simpler reduced-order approximation for design exploration, while recognizing that directional inertial resistance may become important depending on the operating condition.

\section{Influence of thermal dispersion on the optimized design}
\label{app:thermal_dispersion}

This appendix examines the influence of thermal dispersion on the optimized pin-fin arrangement using a correlation identified from unit-cell calculations. The optimizations with and without dispersion use the same stagnant conductivity $k_0(d)$.

\subsection{Unit-cell calculations and separation of heat fluxes}

For each unit-cell RANS solution obtained in Appendix~\ref{app:directional_identification}, two thermal cell problems were solved with a temperature difference $\Delta T=\mathrm{0.1}$~K imposed separately in the $x$ and $y$ directions. The temperature fields were written as
\begin{equation}
T^{(j)}(\mathbf{x})
=T_{\mathrm{ref}}+\frac{\Delta T}{s}x_j+\theta^{(j)}(\mathbf{x}),
\qquad j=x,y,
\label{eq:app_temperature_corrector}
\end{equation}
where $\theta^{(j)}$ is periodic on opposite boundaries of both phases and satisfies $\int_Y\theta^{(j)}\,dY=0$. Temperature and normal diffusive heat flux were continuous at the fluid--solid interface. The governing equations in the fluid and solid regions, $Y_f$ and $Y_s$, respectively, were
\begin{equation}
\begin{aligned}
\rho_f c_{p,f}\mathbf{u}^{\mathrm{cell}}\cdot\nabla T^{(j)}
&=\nabla\cdot\left[\left(k_f+\frac{c_{p,f}\mu_t^{\mathrm{cell}}}{Pr_t}\right)
\nabla T^{(j)}\right]+\dot{q}_{\mathrm{comp}}^{(j)}
&&\text{in }Y_f,\\
0&=\nabla\cdot\left(k_s\nabla T^{(j)}\right)
&&\text{in }Y_s.
\end{aligned}
\label{eq:app_thermal_cell}
\end{equation}
Here, $\mathbf{u}^{\mathrm{cell}}$ and $\mu_t^{\mathrm{cell}}$ are the unit-cell RANS fields. The compensating source $\dot{q}_{\mathrm{comp}}^{(j)}=\rho_f c_{p,f}\langle u_j^{\mathrm{cell}}\rangle_f\Delta T/s$ balances the mean advection, with $\langle\mathbf{u}^{\mathrm{cell}}\rangle_f=\mathbf{u}/\phi$ denoting the intrinsic fluid-average velocity.

The spatial dispersion and turbulent heat-diffusion tensors were evaluated from these temperature fields~\cite{nakayama99} as
\begin{equation}
\begin{aligned}
\left(\mathbf{k}_{\mathrm{dis}}^{\mathrm{cell}}\right)_{ij}
&=-\frac{\rho_f c_{p,f}s}{|Y|\Delta T}
\int_{Y_f}\left(u_i^{\mathrm{cell}}-\langle u_i^{\mathrm{cell}}\rangle_f\right)
\theta^{(j)}\,dY,\\
\left(\mathbf{k}_{\mathrm{turb}}^{\mathrm{cell}}\right)_{ij}
&=\frac{s}{|Y|\Delta T}
\int_{Y_f}\frac{c_{p,f}\mu_t^{\mathrm{cell}}}{Pr_t}
\frac{\partial T^{(j)}}{\partial x_i}\,dY,
\end{aligned}
\label{eq:app_cell_flux_contributions}
\end{equation}
where $i,j=x,y$. The conductive contribution $\mathbf{k}_{\mathrm{cond}}^{\mathrm{cell}}$ was evaluated from $-k_f\nabla T^{(j)}$ in $Y_f$ and $-k_s\nabla T^{(j)}$ in $Y_s$ using the same normalization. The explicit turbulent contribution was subtracted from the total cell conductivity $\mathbf{k}_{\mathrm{tot}}^{\mathrm{cell}}$, defined with the mean advective heat flux excluded:
\begin{equation}
\mathbf{k}_{\mathrm{tot}}^{\mathrm{cell}}
-\mathbf{k}_{\mathrm{turb}}^{\mathrm{cell}}
=\mathbf{k}_{\mathrm{cond}}^{\mathrm{cell}}
+\mathbf{k}_{\mathrm{dis}}^{\mathrm{cell}}.
\label{eq:app_turbulent_flux_subtraction}
\end{equation}
This subtraction was performed in postprocessing, retaining turbulent diffusion in the temperature cell problems. Only $\mathbf{k}_{\mathrm{dis}}^{\mathrm{cell}}$ was used to identify the dispersion correlation.

\subsection{Four-parameter dispersion correlation}

The longitudinal and transverse responses were extracted as $\mathbf{e}_L^{\mathsf{T}}\mathbf{k}_{\mathrm{dis}}^{\mathrm{cell}}\mathbf{e}_L$ and $\mathbf{e}_T^{\mathsf{T}}\mathbf{k}_{\mathrm{dis}}^{\mathrm{cell}}\mathbf{e}_T$, where $\mathbf{e}_L=\mathbf{u}/|\mathbf{u}|$ and $\mathbf{e}_T=(-u_y,u_x)^{\mathsf{T}}/|\mathbf{u}|$. Thus, the projection uses the computed velocity direction, which can differ from the imposed pressure-gradient direction.

The correlation form was based on the high-P\'eclet-number transverse-dispersion correlation of Kuwahara \emph{et al.}~\cite{kuwahara96} for periodic arrays of square rods, which combines linear velocity dependence with a power-law dependence on the solid fraction $1-\phi$. The same functional form was adopted for the longitudinal response, and the amplitude and porosity exponent of each response were independently identified from the present cylindrical-array data. An equally weighted least-squares fit in logarithmic variables was applied to the positive, nonzero-flow responses pooled from all three orientations, yielding
\begin{equation}
\begin{aligned}
k_{\mathrm{dis},L}
&=0.02559(1-\phi)^{-0.86726}\rho_f c_{p,f}\,2r(d)|\mathbf{u}|,\\
k_{\mathrm{dis},T}
&=0.06150(1-\phi)^{0.86703}\rho_f c_{p,f}\,2r(d)|\mathbf{u}|.
\end{aligned}
\label{eq:app_fitted_dispersion}
\end{equation}
The mean absolute relative fitting errors were 23.1\% and 15.3\% for the longitudinal and transverse responses, respectively, using the cell responses as the reference. The diameter-based P\'eclet number $\rho_f c_{p,f}\,2r(d)|\mathbf{u}|/k_f$ ranged from $1.62\times10^4$ to $7.31\times10^5$.

\subsection{Macroscopic model and comparison of optimized designs}

The supplementary macroscopic energy model replaces $k_{\mathrm{eff}}$ in Eq.~\eqref{eq:energy} with
\begin{equation}
\mathbf{k}_{\mathrm{eff}}
=\left[k_0(d)+\phi\frac{c_{p,f}\mu_t}{Pr_t}\right]\mathbf{I}
+\mathbf{k}_{\mathrm{dis}},
\label{eq:app_effective_conductivity_dispersion}
\end{equation}
where
\begin{equation}
\mathbf{k}_{\mathrm{dis}}
=k_{\mathrm{dis},T}\mathbf{I}
+\left(k_{\mathrm{dis},L}-k_{\mathrm{dis},T}\right)
\frac{\mathbf{u}\mathbf{u}^{\mathsf{T}}}{|\mathbf{u}|^2},
\label{eq:app_dispersion_tensor}
\end{equation}
for $|\mathbf{u}|>0$, with $\mathbf{k}_{\mathrm{dis}}=\mathbf{0}$ at zero velocity. The longitudinal direction therefore follows the local Darcy velocity. In this model, the conductive part under flow is approximated by its stagnant value, $\mathbf{k}_{\mathrm{cond}}^{\mathrm{cell}}\simeq k_0(d)\mathbf{I}$. The macroscopic turbulent viscosity $\mu_t$ is supplied by Eq.~\eqref{eq:mut}, and turbulent heat diffusion is represented once through $\phi c_{p,f}\mu_t/Pr_t$.

An additional optimization was performed under the baseline condition in Section~\ref{sec:baseline}. The initial design, flow model, boundary conditions, heat source, porosity constraint, meshes, and optimization settings were unchanged.

Table~\ref{tab:thermal_conductivity_magnitudes} compares the stagnant, turbulent, and dispersive conductivity contributions evaluated using the optimized baseline reduced-model solution.
Here, $k_{\mathrm{turb}}=\phi c_{p,f}\mu_t/Pr_t$, and the dispersion coefficients are evaluated from Eq.~\eqref{eq:app_fitted_dispersion} using the same design variable and Darcy-velocity fields.
In terms of area-averaged values, the turbulent conductivity is approximately 705 times the stagnant conductivity, while the longitudinal and transverse dispersion coefficients are approximately 2.86 and 0.436 times the turbulent conductivity, respectively.

\begin{table}
\centering
\caption{Area-averaged, minimum, and maximum conductivity contributions over the design domain $\Omega$, evaluated using the optimized baseline reduced-model solution.
All values are in W\,m$^{-1}$\,K$^{-1}$.}
\label{tab:thermal_conductivity_magnitudes}
\begin{tabular}{lrrr}
\toprule
Conductivity contribution & Area average & Minimum & Maximum \\
\midrule
$k_0$                 & 1.367 & 0.6733 & 2.157 \\
$k_{\mathrm{turb}}$   & 963.3 & 34.03  & $1.365\times10^4$ \\
$k_{\mathrm{dis},L}$  & 2755  & 0.8143 & $2.982\times10^4$ \\
$k_{\mathrm{dis},T}$  & 419.8 & 0.1694 & 5185 \\
\bottomrule
\end{tabular}
\end{table}

\begin{figure}
\centering
\includegraphics[width=.9\textwidth]{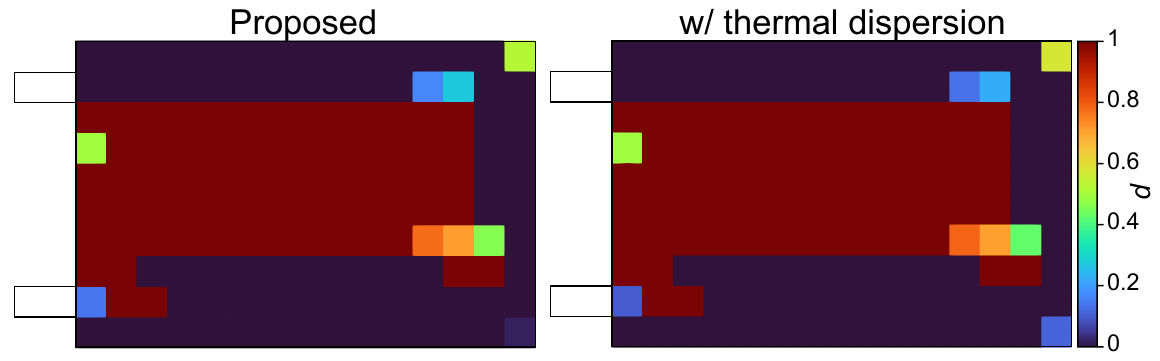}
\caption{Optimized design variable distributions without thermal dispersion (left) and with thermal dispersion (right) under the baseline condition ($U_{\mathrm{in}}=1$~m/s). Both models use the same stagnant conductivity $k_0(d)$. The two panels share the same color scale.}
\label{fig:thermal_dispersion_design}
\end{figure}

The optimized objective values were $J=0.901$~K without thermal dispersion and $J=0.900$~K with thermal dispersion, each evaluated using the thermal model employed in the corresponding optimization. Figure~\ref{fig:thermal_dispersion_design} compares the optimized design variable distributions. Both designs exhibit the same dominant low-resistance paths near the outer walls, with small local differences in the design variable values. Thus, the dominant design pattern was preserved when thermal dispersion was included under the baseline condition examined here. The model without an explicit thermal-dispersion term is therefore retained for the design exploration in the main text.

\section{Sensitivity of the optimization results to fine-mesh resolution}
\label{app:rm_mesh}

The baseline optimization described in Section~3.2 was repeated using three different resolutions of the fine mesh. In the dual-mesh framework, the governing equations of the reduced model are solved and the sensitivities are computed on the fine mesh, whereas the design variables are updated on the coarse mesh. The three fine meshes are denoted RM Mesh~1, RM Mesh~2, and RM Mesh~3 in order of increasing element count. The $15 \times 10$ coarse mesh was kept unchanged, together with the initial design, effective-property correlations, boundary conditions, heat source, constraints, and optimization settings.

Table~\ref{tab:rm_mesh} summarizes the number of elements, the final reduced-model objective value $J_{\mathrm{RM}}$, and the final porosity-constraint value $g$ for each mesh. The objective and constraint values were evaluated on the mesh used in the corresponding optimization.

\begin{table}[tbp]
    \centering
    \caption{Sensitivity of the reduced-model optimization results to fine-mesh resolution under the baseline condition. Relative differences are absolute differences from the RM Mesh~3 results, normalized by those results and expressed as percentages.}
    \label{tab:rm_mesh}
    \begin{tabular}{@{}lrrrrr@{}}
        \toprule
        Mesh & Number of elements & Final $J_{\mathrm{RM}}$ [K] & Final $g$
             & \multicolumn{2}{c}{Relative difference [\%]} \\
        \cmidrule(lr){5-6}
             & & & & $J_{\mathrm{RM}}$ & $g$ \\
        \midrule
        RM Mesh~1 & 4,481 & 0.9016 & 1.0 & 0.09 & 0.00 \\
        RM Mesh~2 & 6,312 & 0.9014 & 1.0 & 0.07 & 0.00 \\
        RM Mesh~3 & 11,505 & 0.9008 & 1.0 & -- & -- \\
        \bottomrule
    \end{tabular}
\end{table}

All three final optimized objective values differed from the RM Mesh~3 result by less than 0.1\%. Increasing the number of elements from 6312 (RM Mesh~2) to 11,505 (RM Mesh~3), an increase of 82.3\%, changed the final optimized objective value by only 0.07\%.

\begin{figure}[pos=htbp]
    \centering
    \IfFileExists{figs/rm_mesh.pdf}{%
        \includegraphics[width=.9\textwidth]{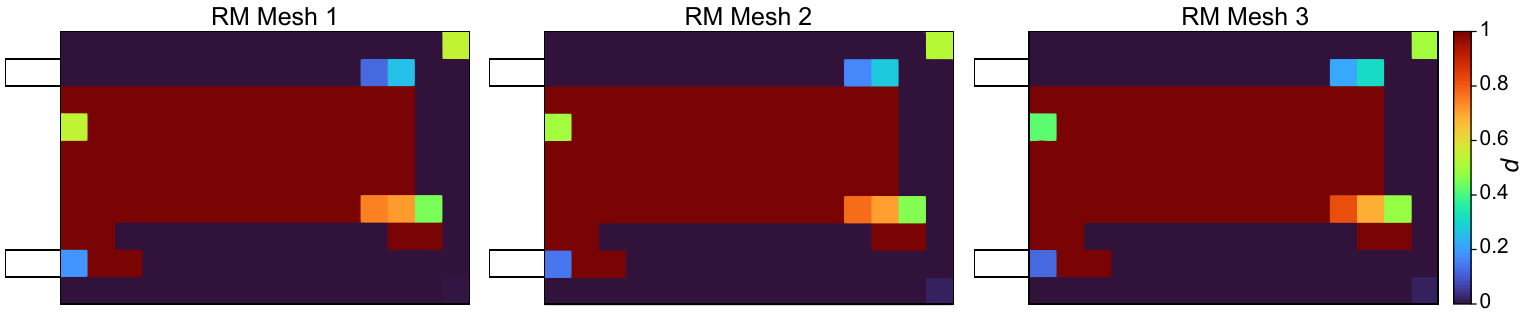}%
    }{%
        \fbox{\parbox[c][30mm][c]{.9\textwidth}{\centering XX}}%
    }
    \caption{Optimized design variable distributions obtained using RM Meshes~1--3 under the baseline condition.}
    \label{fig:rm_mesh}
\end{figure}

Figure~\ref{fig:rm_mesh} compares the optimized design variable distributions obtained using RM Meshes~1--3. The distributions show that all three meshes produce similar results.

Given the small variation in final objective values and the similar optimized design variable distributions over the tested mesh range, RM Mesh~2 was retained for the baseline design exploration.

\section*{Declaration of generative AI and AI-assisted technologies in the manuscript preparation process}
During the preparation of this work, the authors used ChatGPT (OpenAI) for English translation and proofreading and to improve the clarity of the manuscript. After using this tool, the authors reviewed and edited the content as needed and take full responsibility for the content of the published article.

\bibliographystyle{elsarticle-num}
\bibliography{combined_citations_authorYY}

@article{launder74,
title = {The numerical computation of turbulent flows},
journal = {Computer Methods in Applied Mechanics and Engineering},
volume = {3},
number = {2},
pages = {269-289},
year = {1974},
issn = {0045-7825},
doi = {10.1016/0045-7825(74)90029-2},
author = {B. E. Launder and D. B. Spalding},
}

@article{bendse88,
title = {Generating optimal topologies in structural design using a homogenization method},
journal = {Computer Methods in Applied Mechanics and Engineering},
volume = {71},
number = {2},
pages = {197-224},
year = {1988},
issn = {0045-7825},
doi = {10.1016/0045-7825(88)90086-2},
author = {Martin Philip Bends{\o}e and Noboru Kikuchi},
}

@ARTICLE{mudawar00,
  author  = {Mudawar, I.},
  title   = {Assessment of high-heat-flux thermal management schemes},
  journal = {IEEE Transactions on Components and Packaging Technologies},
  year    = {2001},
  volume  = {24},
  number  = {2},
  pages   = {122-141},
  month   = {June},
  doi     = {10.1109/6144.926375},
}

@INPROCEEDINGS{banthiya22,
  author={Banthiya, Abhijeet and Ozguc, Serdar and Pan, Liang and Weibel, Justin A.},
  booktitle={2022 21st IEEE Intersociety Conference on Thermal and Thermomechanical Phenomena in Electronic Systems (iTherm)},
  title={Topology Optimization of an Air-Cooled Heat Sink for Transient Heat Dissipation using a Homogenization Approach},
  year={2022},
  volume={},
  number={},
  pages={1-7},
  doi={10.1109/iTherm54085.2022.9899571}}

@article{saito26,
title = {Multi-scale topology optimization of porous heat sinks with voided lattice structure using a two-layer {Darcy}--{Forchheimer} model},
journal = {International Journal of Heat and Mass Transfer},
volume = {259},
pages = {128324},
year = {2026},
issn = {0017-9310},
doi = {10.1016/j.ijheatmasstransfer.2025.128324},
author = {Tatsuki Saito and Yuto Kikuchi and Kuniharu Ushijima and Kentaro Yaji},
}

@article{subramaniam19,
title = {Topology optimization of conjugate heat transfer systems: A competition between heat transfer enhancement and pressure drop reduction},
journal = {International Journal of Heat and Fluid Flow},
volume = {75},
pages = {165-184},
year = {2019},
issn = {0142-727X},
doi = {10.1016/j.ijheatfluidflow.2019.01.002},
author = {V. Subramaniam and T. Dbouk and J.-L. Harion},
}

@article{togun25,
title = {A comprehensive review of battery thermal management systems for electric vehicles: Enhancing performance, sustainability, and future trends},
journal = {International Journal of Hydrogen Energy},
volume = {97},
pages = {1077-1107},
year = {2025},
issn = {0360-3199},
doi = {10.1016/j.ijhydene.2024.11.093},
author = {Hussein Togun and Ali Basem and Jameel M. dhabab and Hayder I. Mohammed and Abdellatif M. Sadeq and Nirmalendu Biswas and Tuqa Abdulrazzaq and Husam Abdulrasool Hasan and Raad Z. Homod and Pouyan Talebizadehsardari},
}

@article{fawaz22,
title = {Topology optimization of heat exchangers: A review},
journal = {Energy},
volume = {252},
pages = {124053},
year = {2022},
issn = {0360-5442},
doi = {10.1016/j.energy.2022.124053},
author = {Ahmad Fawaz and Yuchao Hua and Steven {Le Corre} and Yilin Fan and Lingai Luo},
}

@article{yeranee21,
title = {A review of recent studies on rotating internal cooling for gas turbine blades},
journal = {Chinese Journal of Aeronautics},
volume = {34},
number = {7},
pages = {85-113},
year = {2021},
issn = {1000-9361},
doi = {10.1016/j.cja.2020.12.035},
author = {Kirttayoth Yeranee and Yu Rao},
}

@article{marshall19,
title = {Thermal Management of Vehicle Cabins, External Surfaces, and Onboard Electronics: An Overview},
journal = {Engineering},
volume = {5},
number = {5},
pages = {954-969},
year = {2019},
issn = {2095-8099},
doi = {10.1016/j.eng.2019.02.009},
author = {Garrett J. Marshall and Colin P. Mahony and Matthew J. Rhodes and Steve R. Daniewicz and Nicholas Tsolas and Scott M. Thompson},
}

@article{takezawa19,
title = {Method to optimize an additively-manufactured functionally-graded lattice structure for effective liquid cooling},
journal = {Additive Manufacturing},
volume = {28},
pages = {285-298},
year = {2019},
issn = {2214-8604},
doi = {10.1016/j.addma.2019.04.004},
author = {Akihiro Takezawa and Xiaopeng Zhang and Masaki Kato and Mitsuru Kitamura},
}

@article{takezawa24,
title = {Validity of the quasi-{2D} optimal variable density lattice for effective liquid cooling based on {Darcy}--{Forchheimer} theory},
journal = {Thermal Science and Engineering Progress},
volume = {55},
pages = {102898},
year = {2024},
issn = {2451-9049},
doi = {10.1016/j.tsep.2024.102898},
author = {Akihiro Takezawa and Kenjiro Matsui and Shomu Murakoshi and Kentaro Taniguchi and Ryota Moritoyo and Mitsuru Kitamura},
}

@Article{alexandersen20,
AUTHOR = {Alexandersen, Joe and Andreasen, Casper Schousboe},
TITLE = {A Review of Topology Optimisation for Fluid-Based Problems},
JOURNAL = {Fluids},
VOLUME = {5},
YEAR = {2020},
NUMBER = {1},
pages = {29},
ISSN = {2311-5521},
DOI = {10.3390/fluids5010029}
}

@Article{sun23,
AUTHOR = {Sun, Yiwei and Hao, Menglong and Wang, Zexu},
TITLE = {Topology Optimization of Turbulent Flow Cooling Structures Based on the {$k$-$\varepsilon$} Model},
JOURNAL = {Entropy},
VOLUME = {25},
YEAR = {2023},
NUMBER = {9},
pages = {1299},
PubMedID = {37761598},
ISSN = {1099-4300},
DOI = {10.3390/e25091299}
}

@article{sparrow80,
    author = {Sparrow, E. M. and Ramsey, J. W. and Altemani, C. A. C.},
    title = {Experiments on In-line Pin Fin Arrays and Performance Comparisons with Staggered Arrays},
    journal = {Journal of Heat Transfer},
    volume = {102},
    number = {1},
    pages = {44-50},
    year = {1980},
    month = {02},
    issn = {0022-1481},
    doi = {10.1115/1.3244247},
}

@article{vanfossen82,
    author = {VanFossen, G. J.},
    title = {Heat-Transfer Coefficients for Staggered Arrays of Short Pin Fins},
    journal = {Journal of Engineering for Power},
    volume = {104},
    number = {2},
    pages = {268-274},
    year = {1982},
    month = {04},
    issn = {0022-0825},
    doi = {10.1115/1.3227275},
}

@article{metzger82,
    author = {Metzger, D. E. and Berry, R. A. and Bronson, J. P.},
    title = {Developing Heat Transfer in Rectangular Ducts With Staggered Arrays of Short Pin Fins},
    journal = {Journal of Heat Transfer},
    volume = {104},
    number = {4},
    pages = {700-706},
    year = {1982},
    month = {11},
    issn = {0022-1481},
    doi = {10.1115/1.3245188},
}

@article{metzger84,
    author = {Metzger, D. E. and Fan, C. S. and Haley, S. W.},
    title = {Effects of Pin Shape and Array Orientation on Heat Transfer and Pressure Loss in Pin Fin Arrays},
    journal = {Journal of Engineering for Gas Turbines and Power},
    volume = {106},
    number = {1},
    pages = {252-257},
    year = {1984},
    month = {01},
    issn = {0742-4795},
    doi = {10.1115/1.3239545},
}

@article{armstrong88,
    author = {Armstrong, J. and Winstanley, D.},
    title = {A Review of Staggered Array Pin Fin Heat Transfer for Turbine Cooling Applications},
    journal = {Journal of Turbomachinery},
    volume = {110},
    number = {1},
    pages = {94-103},
    year = {1988},
    month = {01},
    issn = {0889-504X},
    doi = {10.1115/1.3262173},
}

@article{nakayama08,
    author = {Nakayama, A. and Kuwahara, F.},
    title = {A General Macroscopic Turbulence Model for Flows in Packed Beds, Channels, Pipes, and Rod Bundles},
    journal = {Journal of Fluids Engineering},
    volume = {130},
    number = {10},
    pages = {101205},
    year = {2008},
    month = {09},
    issn = {0098-2202},
    doi = {10.1115/1.2969461},
}

@article{dilgen18,
  author    = {Dilgen, Sumer B. and Dilgen, Cetin B. and Fuhrman, David R. and Sigmund, Ole and Lazarov, Boyan S.},
  title     = {Density based topology optimization of turbulent flow heat transfer systems},
  journal   = {Structural and Multidisciplinary Optimization},
  year      = {2018},
  volume    = {57},
  number    = {5},
  pages     = {1905--1918},
  doi       = {10.1007/s00158-018-1967-6},
  publisher = {Springer},
}

@article{kobayashi21,
  author    = {Kobayashi, Hiroki and Yaji, Kentaro and Yamasaki, Shintaro and Fujita, Kikuo},
  title     = {Topology design of two-fluid heat exchange},
  journal   = {Structural and Multidisciplinary Optimization},
  year      = {2021},
  volume    = {63},
  pages     = {821--834},
  doi       = {10.1007/s00158-020-02736-8},
  publisher = {Springer},
}

@article{kuwahara98,
  author    = {Kuwahara, Fujio and Kameyama, Y. and Yamashita, S. and Nakayama, Akira},
  title     = {Numerical Modeling of Turbulent Flow in Porous Media Using a Spatially Periodic Array},
  journal   = {Journal of Porous Media},
  year      = {1998},
  volume    = {1},
  number    = {1},
  pages     = {47--55},
  doi       = {10.1615/JPorMedia.v1.i1.40},
  publisher = {Begell House}
}

@article{padhy24,
  author    = {Padhy, Rahul Kumar and Chandrasekhar, Aaditya and Suresh, Krishnan},
  title     = {{FluTO}: Graded multi-scale topology optimization of large contact area fluid-flow devices using neural networks},
  journal   = {Engineering with Computers},
  year      = {2024},
  volume    = {40},
  pages     = {971--987},
  doi       = {10.1007/s00366-023-01827-6},
  publisher = {Springer},
}

@article{borrvallxx,
author = {Borrvall, Thomas and Petersson, Joakim},
title = {Topology optimization of fluids in {Stokes} flow},
journal = {International Journal for Numerical Methods in Fluids},
volume = {41},
number = {1},
pages = {77-107},
doi = {10.1002/fld.426},
year = {2003}
}

@article{svanbergxx,
author = {Svanberg, Krister},
title = {The method of moving asymptotes---a new method for structural optimization},
journal = {International Journal for Numerical Methods in Engineering},
volume = {24},
number = {2},
pages = {359-373},
doi = {10.1002/nme.1620240207},
year = {1987}
}

@article{pietropaoli19,
  author    = {Pietropaoli, M. and Montomoli, F. and Gaymann, A.},
  title     = {Three-dimensional fluid topology optimization for heat transfer},
  journal   = {Structural and Multidisciplinary Optimization},
  year      = {2019},
  volume    = {59},
  pages     = {801--812},
  doi       = {10.1007/s00158-018-2102-4},
  publisher = {Springer},
}

@article{thillaithevan25,
  author       = {Thillaithevan, Dilaksan and Hewson, Robert and Murphy, Ryan and Santer, Matthew and Carver, Alex and Nikiteas, Giannis and Raske, Nicholas},
  title        = {A subspace method for {3D} multiscale heat sink modelling and optimization},
  journal      = {Structural and Multidisciplinary Optimization},
  year         = {2025},
  volume       = {68},
  pages        = {167},
  doi          = {10.1007/s00158-025-04088-7},
  publisher    = {Springer},
}

@article{yoon10,
  author    = {Gil Ho Yoon},
  title     = {Topological design of heat dissipating structure with forced convective heat transfer},
  journal   = {Journal of Mechanical Science and Technology},
  year      = {2010},
  volume    = {24},
  number    = {6},
  pages     = {1225--1233},
  doi       = {10.1007/s12206-010-0328-1},
  publisher = {Springer}
}

@article{zhao21,
  author    = {Zhao, Jiaqi and Zhang, Ming and Zhu, Yu and Cheng, Rong and Wang, Leijie},
  title     = {Topology optimization of turbulent forced convective heat sinks using a multi-layer thermofluid model},
  journal   = {Structural and Multidisciplinary Optimization},
  year      = {2021},
  volume    = {64},
  pages     = {3835--3859},
  doi       = {10.1007/s00158-021-03064-1},
  publisher = {Springer},
}

@book{nield17,
  author    = {Nield, Donald A. and Bejan, Adrian},
  title     = {Convection in Porous Media},
  edition   = {5},
  year      = {2017},
  publisher = {Springer},
  address   = {Cham},
  isbn      = {978-3-319-49561-3; 978-3-319-49562-0},
  doi       = {10.1007/978-3-319-49562-0},
}

@article{hughes89,
  author  = {Hughes, Thomas J. R. and Franca, Leopoldo P. and Hulbert, Gregory M.},
  title   = {A new finite element formulation for computational fluid dynamics: {VIII}. {The Galerkin}/least-squares method for advective-diffusive equations},
  journal = {Computer Methods in Applied Mechanics and Engineering},
  year    = {1989},
  volume  = {73},
  number  = {2},
  pages   = {173--189},
  doi     = {10.1016/0045-7825(89)90111-4},
}

@article{kuwahara96,
  author  = {Kuwahara, F. and Nakayama, A. and Koyama, H.},
  title   = {A Numerical Study of Thermal Dispersion in Porous Media},
  journal = {Journal of Heat Transfer},
  year    = {1996},
  volume  = {118},
  number  = {3},
  pages   = {756--761},
  doi     = {10.1115/1.2822696},
}

@article{nakayama99,
  author  = {Nakayama, A. and Kuwahara, F.},
  title   = {A Macroscopic Turbulence Model for Flow in a Porous Medium},
  journal = {Journal of Fluids Engineering},
  year    = {1999},
  volume  = {121},
  number  = {2},
  pages   = {427--433},
  doi     = {10.1115/1.2822227},
}

@article{kikuchi26,
  author  = {Yuto Kikuchi and Kentaro Yaji and Kikuo Fujita and Tatsuki Saito and Kuniharu Ushijima},
  title   = {Homogenization-based optimal design of non-uniform lattice heat sinks: Comparative study on cell structures},
  journal = {Case Studies in Thermal Engineering},
  year    = {2026},
  volume  = {81},
  pages   = {108029},
  doi     = {10.1016/j.csite.2026.108029},
}

@article{ohtani26,
  author  = {Kaito Ohtani and Hiroki Kawabe and Kentaro Yaji and Kikuo Fujita and Vikrant Aute},
  title   = {Homogenization-based optimization of wall thickness distribution for {TPMS} two-fluid heat exchangers},
  journal = {International Journal of Heat and Mass Transfer},
  year    = {2026},
  volume  = {267},
  pages   = {128977},
  doi     = {10.1016/j.ijheatmasstransfer.2026.128977},
}

@article{yanagihara26,
  author  = {Kazutaka Yanagihara and Jun Iwasaki and Kiyoto Saso and Taichi Yamashita and Shomu Murakoshi and Akihiro Takezawa},
  title   = {Flow-priority optimization of additively manufactured variable-{TPMS} lattice heat exchanger based on macroscopic analysis},
  journal = {Additive Manufacturing},
  year    = {2026},
  volume  = {125},
  pages   = {105246},
  doi     = {10.1016/j.addma.2026.105246},
}

@article{bayat26,
  author  = {Amirhossein Bayat and Hao Li and Joe Alexandersen},
  title   = {Density-based topology optimization for turbulent fluid flow using the standard $k$--$\varepsilon$ {RANS} model with wall functions imposed through an implicit wall penalty formulation},
  journal = {Computer Methods in Applied Mechanics and Engineering},
  year    = {2026},
  volume  = {460},
  pages   = {119100},
  issn    = {0045-7825},
  doi     = {10.1016/j.cma.2026.119100},
}

@article{ghosh24,
  author  = {Shinjan Ghosh and Sudeepta Mondal and Jayanta S. Kapat and Asok Ray},
  title   = {Parametric shape optimization of pin fin arrays using a multi-fidelity surrogate model based {Bayesian} method},
  journal = {Applied Thermal Engineering},
  year    = {2024},
  volume  = {247},
  pages   = {122876},
  doi     = {10.1016/j.applthermaleng.2024.122876},
}

@article{zhao26,
  author  = {Chunyan Zhao and Guilian Wang and Hongxin Deng and Yunran Wang},
  title   = {Multi-objective optimization of heat sinks with micro variable-density pin-fins for hotspot thermal management},
  journal = {International Journal of Heat and Fluid Flow},
  year    = {2026},
  volume  = {117},
  pages   = {110079},
  doi     = {10.1016/j.ijheatfluidflow.2025.110079},
}

@article{mihalko25,
  author  = {Evan M. Mihalko and Amrita Basak},
  title   = {Harnessing transfer learning for achieving superior thermal-hydraulic performance in heterogeneous pin-fin arrays},
  journal = {International Communications in Heat and Mass Transfer},
  year    = {2025},
  volume  = {165},
  pages   = {108968},
  doi     = {10.1016/j.icheatmasstransfer.2025.108968},
}

\end{document}